\documentclass[acmsmall, screen]{acmart}
\AtBeginDocument{%
  }

\setcopyright{none}

\usepackage{adjustbox}
\usepackage{caption}
\usepackage{graphicx}
\usepackage{pgf-pie}
\usepackage{pifont}
\usepackage{threeparttable}
\usepackage{hyperref}
\usepackage{makecell}
\usepackage{listings}
\usepackage{xcolor}
\usepackage{ulem}
\usepackage[most]{tcolorbox}

\begin{document}

\title{LLMs in Software Security: A Survey of Vulnerability Detection Techniques and Insights}



\author{Ze Sheng}
\affiliation{%
  \institution{Texas A\&M University}
  \city{College Station}
  \country{USA}}
\email{zesheng@tamu.edu}

\author{Zhicheng Chen}
\affiliation{%
  \institution{Texas A\&M University}
  \city{College Station}
  \country{USA}}
\email{zhicheng@tamu.edu}

\author{Shuning Gu}
\affiliation{%
  \institution{Texas A\&M University}
  \city{College Station}
  \country{USA}}
\email{shuning@tamu.edu}

\author{Heqing Huang}
\affiliation{%
  \institution{City University of Hong Kong}
  \city{Hong Kong}
  \country{China}}
\email{heqhuang@cityu.edu.hk}

\author{Guofei Gu}
\affiliation{%
  \institution{Texas A\&M University}
  \city{College Station}
  \country{USA}}
\email{guofei@tamu.edu}

\author{Jeff Huang}
\affiliation{%
  \institution{Texas A\&M University}
  \city{College Station}
  \country{USA}}
\email{jeffhuang@tamu.edu}

\renewcommand{\shortauthors}{Sheng et al.}

\begin{abstract}
Large Language Models (LLMs) are emerging as transformative tools for software vulnerability detection. Traditional methods, including static and dynamic analysis, face limitations in efficiency, false-positive rates, and scalability with modern software complexity. Through code structure analysis, pattern identification, and repair suggestion generation, LLMs demonstrate a novel approach to vulnerability mitigation.

This survey examines LLMs in vulnerability detection, analyzing problem formulation, model selection, application methodologies, datasets, and evaluation metrics. We investigate current research challenges, emphasizing cross-language detection, multimodal integration, and repository-level analysis. Based on our findings, we propose solutions addressing dataset scalability, model interpretability, and low-resource scenarios.

Our contributions include: (1) a systematic analysis of LLM applications in vulnerability detection; (2) a unified framework examining patterns and variations across studies; and (3) identification of key challenges and research directions. This work advances the understanding of LLM-based vulnerability detection. The latest findings are maintained at \href{https://github.com/OwenSanzas/LLM-For-Vulnerability-Detection}{https://github.com/OwenSanzas/LLM-For-Vulnerability-Detection}
\end{abstract}



\keywords{Large Language Models, Vulnerability Detection, Cybersecurity}


\maketitle

\section{Introduction}
\begin{figure}[h]
  \includegraphics[width=0.6\linewidth]{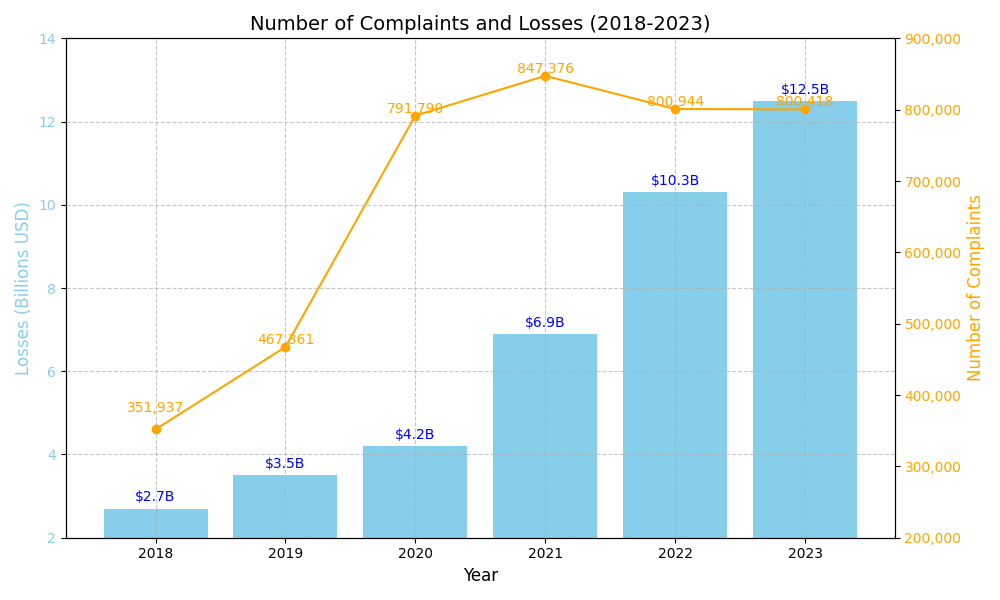}
  \caption{Complaints and Financial Losses from 2018-2023}
  \Description{Complaint_Loss}
  \label{fig:complaint_loss}
\end{figure}

Vulnerability detection plays an important part in the design and maintenance of modern software. Statistical evidence indicates that approximately 70\% of security vulnerabilities originate from defects in the software development process \cite{Aslan2023}. According to the metrics provided by Common Vulnerabilities and Exposures Numbering Authorities (CNAs), a growth is witnessed that in the past 5 years, about 120,000 CVEs have been discovered and reported \cite{CWE_Metrics}. According to FBI's cybercrime report shown in Figure \ref{fig:complaint_loss}, the period from 2018 to 2023 suffers from a large amount of cybersecurity crimes and complaints. A recent example is the CrowdStrike incident in July 2024 \cite{GAO2024}, where a faulty software update caused widespread system crashes across critical infrastructure sectors including healthcare, transportation, and finance. Therefore, enhanced focus and investment in vulnerability detection technology is in demand.

State-of-the-art vulnerability detection approaches/tools can be broadly classified into static analysis and dynamic analysis \cite{chess2004static, livshits2005finding, zaddach2014avatar, russo2010dynamic}. Static analysis examines source code or bytecode to identify potential security vulnerabilities while dynamic analysis does it during program execution, including techniques such as fuzz testing \cite{Serebryany2016}. Fuzz testing identifies potential vulnerabilities by inputting random or specific invalid data into applications and observing system responses. However, with the increasing scale of software systems, both traditional static and dynamic analysis approaches have distinct limitations. For example, static and dynamic analysis tools suffer from high false positive rates, low efficiency and vast effort for overwhelmed amount of types of vulnerabilities \cite{Johnson13, Venkatasubramanyam14}.

Large Language Models (LLMs), as an advancement in natural language processing (NLP), are trained by deep learning techniques, particularly the Transformer architecture to focus on diverse NLP tasks \cite{vaswani2023attentionneed}. Recently, LLMs have shown remarkable abilities in software development area \cite{Hou24}. Based on these capabilities, several LLM-driven vulnerability detection methodologies and tools have been proposed by researchers. This new trend has attracted attention from both cybersecurity experts and machine learning researchers, potentially revolutionizing this field. For example, in 2024 DARPA held the Artificial Intelligence Cyber Challenge (AIxCC), where competitors only leverage general-purpose LLMs (GPT series, Claude series, and Gemini series) for vulnerability detection, reproduction, and patching ~\cite{AIxCC}. The competition brought together leading experts in both cybersecurity and machine learning all over the world, indicating the significant potential of applying LLMs to vulnerability detection.


The increasing adoption of LLMs in vulnerability detection is evident from initiatives like AIxCC, showcasing the rapid evolution of their capabilities and the diversity of approaches in this field. Despite growing interest and significant research efforts, a systematic survey that thoroughly examines LLM-based detection methods is still lacking.

This paper addresses this gap by presenting the first comprehensive survey focused on understanding the strengths and weaknesses of LLMs in vulnerability detection. To provide a structured and holistic analysis, we construct our survey around four key aspects: (1) identifying effective LLM architectures for security tasks, (2) evaluating benchmarks, datasets, and metrics for reliable assessment, (3) analyzing techniques to uncover best practices, and (4) recognizing challenges to guide future research directions. These aspects collectively highlight the current state of the field and its potential for advancement.

Based on these aspects, we focus on the following research questions (RQs):

\begin{itemize}
\item \textbf{RQ1}. What LLMs have been applied to vulnerability detection?
\item \textbf{RQ2}. What benchmarks, dataset and metrics have been designed to evaluate vulnerability detection?
\item \textbf{RQ3}. What techniques have been used in LLM for vulnerability detection?
\item \textbf{RQ4}. What are the challenges that LLMs are facing in detecting vulnerabilities and potential directions to solve them?

\end{itemize}

To analyze and summarize the RQs, we selected more than 80 papers (with 58 highly related papers) from over 500 papers to ensure the recency (2019-2024) and the relevance (focusing on LLMs in vulnerability detection).  Therefore, this survey will not discuss traditional machine learning approaches (conventional CNN, RNN and LSTM) in vulnerability detection, which were predominantly used between 2014 and 2020. 

In general, our key findings can be summarized as the positive and key gaps:

\begin{itemize}
    \item \textbf{The Positive: }Cybersecurity community has experienced a positive impact from LLM, evidenced by a substantial increase in published articles in recent years. The contributions span multiple areas, including vulnerability localization, detection and analysis. C, Java and Solidity have emerged as predominant focuses in this domain. Research methodologies consistently emphasize three key components: LLM implementation, prompt engineering, and semantic processing methods. Moreover, multi-agent approaches are widely used because of the decomposition of complex vulnerability detection challenges into manageable sub-problems.

    \item \textbf{Key Gaps:}
        \begin{itemize}
            \item \textbf{Narrow Scope and Limited Repository-Level Coverage:} Current work often restricts itself to binary classification of function-level vulnerabilities within small, specialized datasets. Moreover, few studies address the detection and reproduction of vulnerabilities at the repository level, where cross-file dependencies and longer call stacks pose significant challenges.

            \item \textbf{Rapid Advances in Frontier LLMs:} Breakthroughs in 2023--2024 indicate that fine-tuning and leveraging frontier LLMs will be critical for future progress, yet most research to date relies on less-capable traditional models.

            \item \textbf{Insufficient Context Awareness:} There is insufficient attention to complex, multi-file dependencies and long call stacks. Although neuro-Symbolic approaches (e.g. CodeQL, Bear with LLMs) have shown promise on large-scale projects, improved taint propagation modeling, more efficient LLM reasoning, and cross-language adaptation are still required.

            \item \textbf{Vulnerability Type Imbalance:} Memory-related vulnerabilities (e.g. buffer overflow issues) receive disproportionately higher detection accuracy, while logical vulnerabilities remain relatively underexplored.

            \item \textbf{Dataset Limitations:} Existing datasets are narrowly scoped and have a data leakage problem. A dedicated and comprehensive data set specifically tailored for LLM-based vulnerability detection is urgently needed to drive both fundamental and applied research.

        \end{itemize}
    
\end{itemize}

\begin{figure*}
  \includegraphics[width=0.8\textwidth]{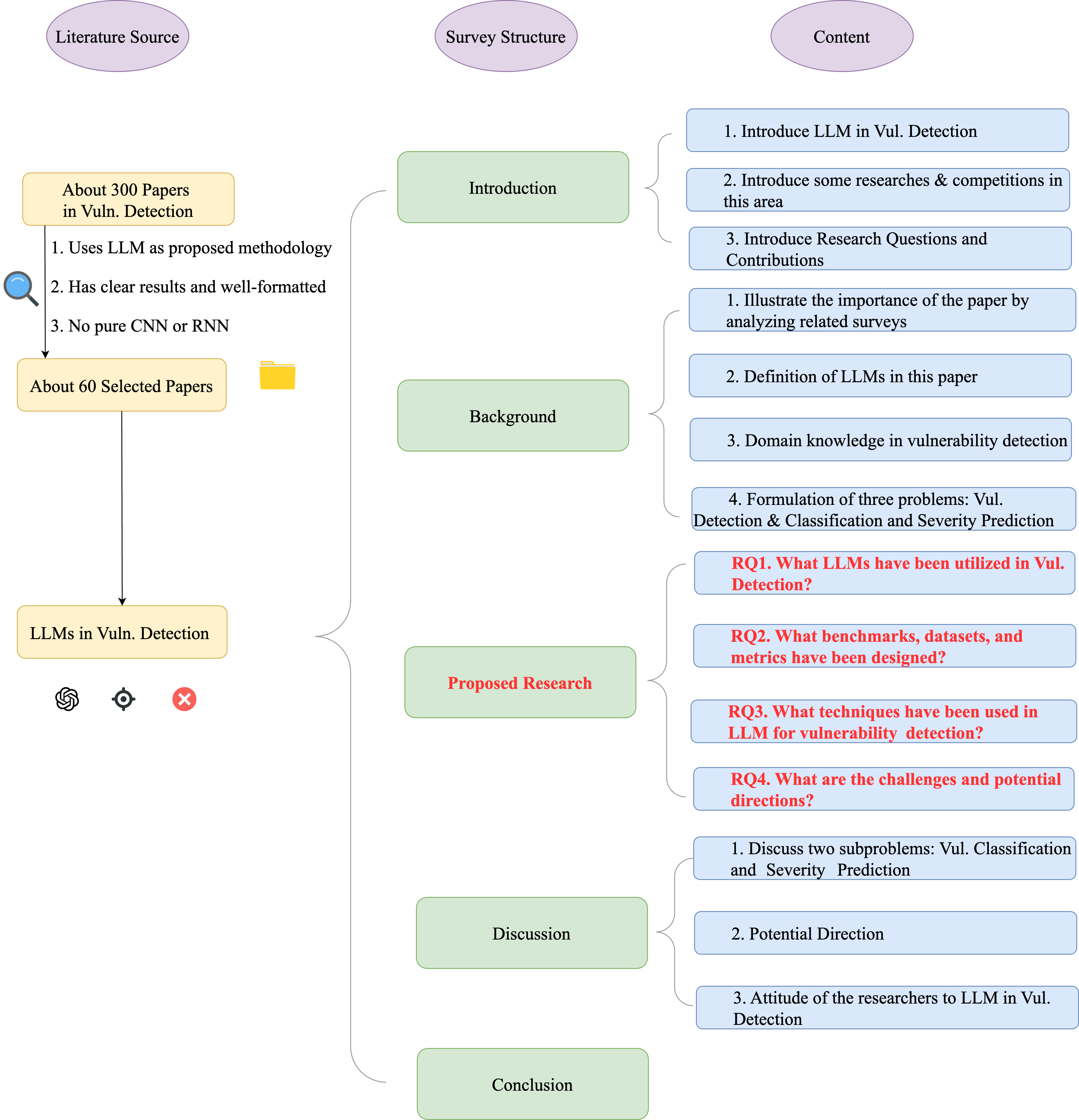}
  \caption{Survey Structure \& Research Selection}
  \Description{Survey Structure \& Research Selection}
  \label{fig:structure}
\end{figure*}

The following sections present a comprehensive analysis of LLMs in vulnerability detection. Given the extensive scope of this survey, this section outlines the structure, main themes, and narrative flow of our analysis. A visualization of this paper's structure is shown in Figure \ref{fig:structure}.

\section{Background}
\subsection{Paper Selection And Scope}
To ensure a comprehensive and systematic review, we began our search with top-tier security conferences, such as IEEE Symposium on Security and Privacy (S\&P), USENIX Security, and ACM Conference on Computer and Communications Security (CCS), as well as journals like IEEE Transactions on Software Engineering. Then we searched by extracting key terms such as "vulnerability detection," "LLM," "large language model," and "AI" from papers published in conferences and journals. Using these keywords, we conducted iterative searches every three weeks, refining the selection over time. Over a two-month period, we screened approximately 500-600 papers and selected 58 highly relevant studies.

This survey focuses exclusively on the application of LLMs in vulnerability detection, analyzing techniques, datasets, benchmarks, and challenges. We reviewed works targeting programming languages like C/C++, Java, and Solidity, which are the primary focus areas for LLM-based vulnerability detection. Studies centered on traditional machine learning methods, such as CNNs and RNNs, and those unrelated to vulnerability detection, such as malware analysis or network intrusion detection, were excluded.

In terms of datasets, we primarily evaluated function-level and file-level granularity, noting that C/C++ datasets dominate the field. However, repository-level datasets that better reflect real-world development scenarios are significantly lacking. This limitation poses challenges for LLMs in generalizing to complex, multi-file vulnerabilities.

By clearly defining the scope and adopting a rigorous, keyword-driven selection process, we aim to ensure the robustness and relevance of this survey while laying a solid foundation for future research.

\subsection{Related Reviews}
In the past five years, many studies have been proposed on leveraging LLMs for vulnerability detection. Several comprehensive surveys have been presented on vulnerability detection techniques, covering traditional approaches (static/dynamic analysis) and machine learning methods (CNN, RNN) \cite{Aslan2023, Chakraborty2022reveal, Zhang21, eberendu2022systematic, zhu2023application}. They do not specifically address the integration of LLMs in vulnerability detection. Yao et al. \cite{Yao24} reviewed LLMs in the security and privacy domain, proposing their positive impacts, potential threats, and inherent threat. However, their analysis focuses on a broad overview of these issues rather than providing a methodological summary of LLM-based detection approaches. Xu et al. \cite{xu2024large} presented an overview of LLMs in the entire cybersecutiry domain, including malware analysis, network intrusion detection, etc. Our survey specifically focuses on LLM-based vulnerability detection with a more detailed summary of techniques and methodologies. At the same time, Zhou et al. \cite{zhou2024large} investigated how LLMs are adapted for vulnerability detection and repair. While their work provides valuable insights, our survey differs in several aspects: (1) by the time of writing, both OpenAI and Anthropic have released more powerful LLMs (GPT-4o, o1 and Claude 3.5 Sonnet) that have stronger inference abilities and larger context windows; (2) we conduct a comprehensive analysis of benchmarks and evaluation metrics for LLM-based vulnerability detection systems; (3) we focus more on the details of vulnerability detection and understanding.

\subsection{Large Language Models (LLMs)}
Large Language Models (LLMs) have emerged as a significant progress in the evolution of language models \cite{zhao2023survey}. The Transformer architecture has enabled unprecedented scaling capabilities. LLMs are characterized by their massive scale, typically incorporating hundreds of billions of parameters trained on vast corpus. Therefore, it leads to remarkable capabilities in general human tasks \cite{dai2023auggptleveragingchatgpttext}.

\subsection{Vulnerability Detection Problem}
\subsubsection{Domain Knowledge}
Some popular vulnerability databases, such as Common Weakness Enumeration (CWE)\footnote{\url{https://cwe.mitre.org/}}, Common Vulnerabilities and Exposures (CVE)\footnote{\url{https://www.cve.org/}}, Common Vulnerability Scoring System (CVSS)\footnote{\url{https://www.first.org/cvss/}} and National Vulnerability Database (NVD)\footnote{\url{https://nvd.nist.gov/}}, have been built to record the definition and evaluation of common vulnerabilities. CWE focuses on all vulnerabilities in the software development lifecycle (from development to maintenance.) Rather than focusing on specific real-world security vulnerabilities (e.g., Heartbleed and Log4Shell), CWE focuses on the root causes of these real-world vulnerabilities like Use-After-Free (CWE-416) and Out-of-bounds Write (CWE-787). CVE is a public community that identifies and catalogs security vulnerabilities in software and hardware. The community will allocate a unique identifier to each real-world vulnerability. For example, the identifier of Log4Shell is CVE-2021-44228. CVSS is a standardized framework for assessing the risk level of a vulnerability through a number of metrics: exploitability, impact, exploit code maturity, and remediation level, etc. These metrics result in an overall score on a scale of 0 to 10, and severity from low to critical. Log4Shell was given a CVSS score of 10 (severity: critical). NVD is a database that contains basic information about real-world vulnerabilities, such as CVE identifier, technical details of the vulnerability, CVSS score, and mitigation recommendations. For instance, NVD records that Log4Shell can be exploited to remotely execute code on the victim server by constructing malicious JNDI statements,  which is caused by Improper Input Validation (CWE-20) and Uncontrolled Resource Consumption (CWE-400).

\subsubsection{Vulnerability Detection}


\begin{figure}[H]
  \includegraphics[width=1\linewidth]{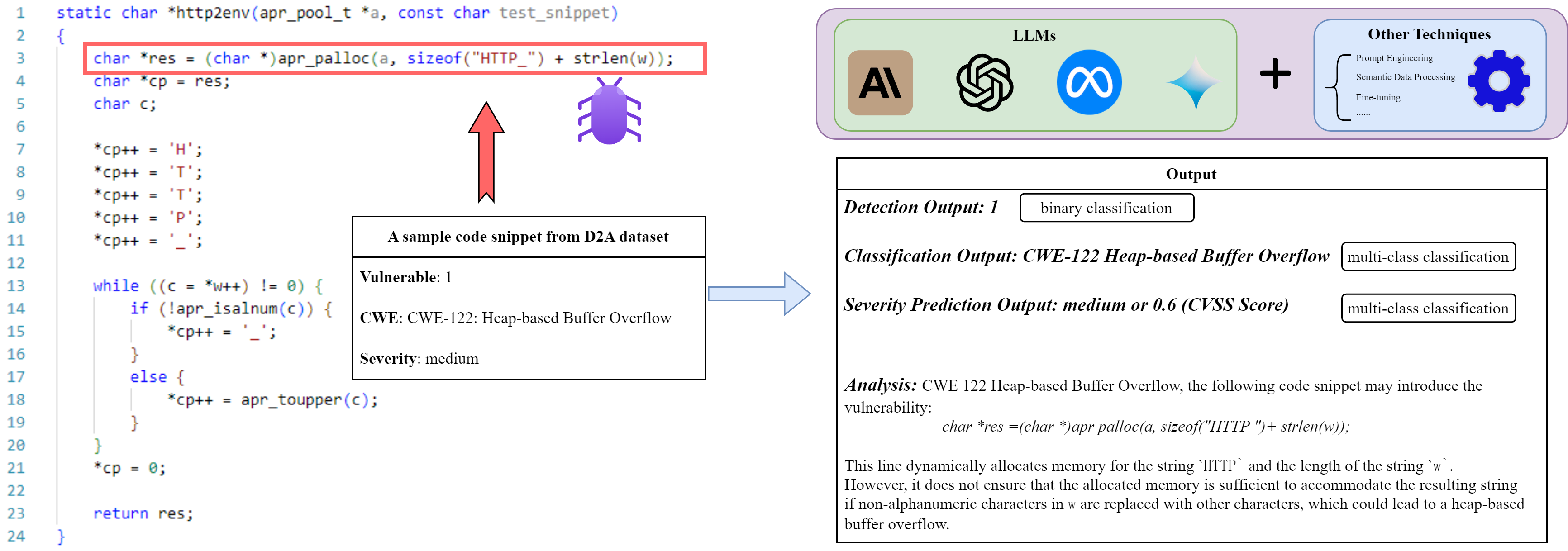}
  \caption{An Example Workflow of Vuln. Detection, Classification and Severity Prediction}
  \Description{Sample problem}
  \label{fig:Example_Problem}
\end{figure}

Vulnerability detection serves as the core focus of all selected papers in this survey, representing the primary application of LLMs in software security. The fundamental task can be formally defined as a binary classification problem:

Let $C_i$ denote the input source code and $VD_i$ represent an LLM-driven vulnerability detector. The output $Y_i \in \{0,1\}$ indicates the vulnerability status where $Y_i = 1$ indicates that the code is vulnerable, and $Y_i = 0$ indicates that the code is non-vulnerable.

Figure \ref{fig:Example_Problem} shows an example workflow of vulnerability detection in most of selected researches, and two subproblems: vulnerability classification and severity prediction.

\subsubsection{Vulnerability Classification}
Beyond binary detection, some studies explore LLMs' capability in multi-class vulnerability classification to enhance model reliability. This task requires LLMs to not only identify the presence of vulnerabilities but also determine their specific types according to established standards like CWE.

Formally, vulnerability classification can be defined as: Let $C_i$ denote the input source code and $VC_i$ represent an LLM-driven vulnerability classifier. The output $Y_i$ indicates the specific vulnerability type:
$Y_i = VC_i(C_i) \in \{type_1, type_2, ..., type_n\}$, where $type_i$ could be vulnerability names (e.g., Buffer Overflow, SQL Injection) or standardized identifiers (e.g., CWE-119, CWE-89).

For example, when examining a code snippet, an LLM might not only detect its vulnerability but also classify it as "CWE-79: Cross-site Scripting (XSS)", providing more detailed guidance for security mitigation.

\subsubsection{Vulnerability Severity Prediction}
Some studies extend vulnerability analysis to include severity prediction alongside detection. This can be formulated as either a multi-class classification problem or a regression task, depending on the granularity of severity measurement.
Formally, let $C_i$ denote the input source code and $VS_i$ represent an LLM-driven severity predictor. The output \( Y_i \) can be defined in two forms: it may represent a severity score such as "low," "medium," or "high," based on the vulnerability severity level of the input source code \( C_i \).
Different studies have adopted varying approaches to severity prediction:
\begin{itemize}
\item Categorical Classification: Alam et al. \cite{alam2024detection} employ a three-level classification system (high, medium, low) for straightforward severity assessment.
\item Score Prediction: Fu et al.\cite{fu_chatgpt_2023} requires LLMs to predict numerical CVSS scores on a scale of 0 to 10, providing more precise severity measurements.
\end{itemize}
This additional severity information helps prioritize vulnerability remediation efforts and allocate security resources more effectively in practical applications.

\begin{figure*}[h]
  \includegraphics[width=\linewidth]{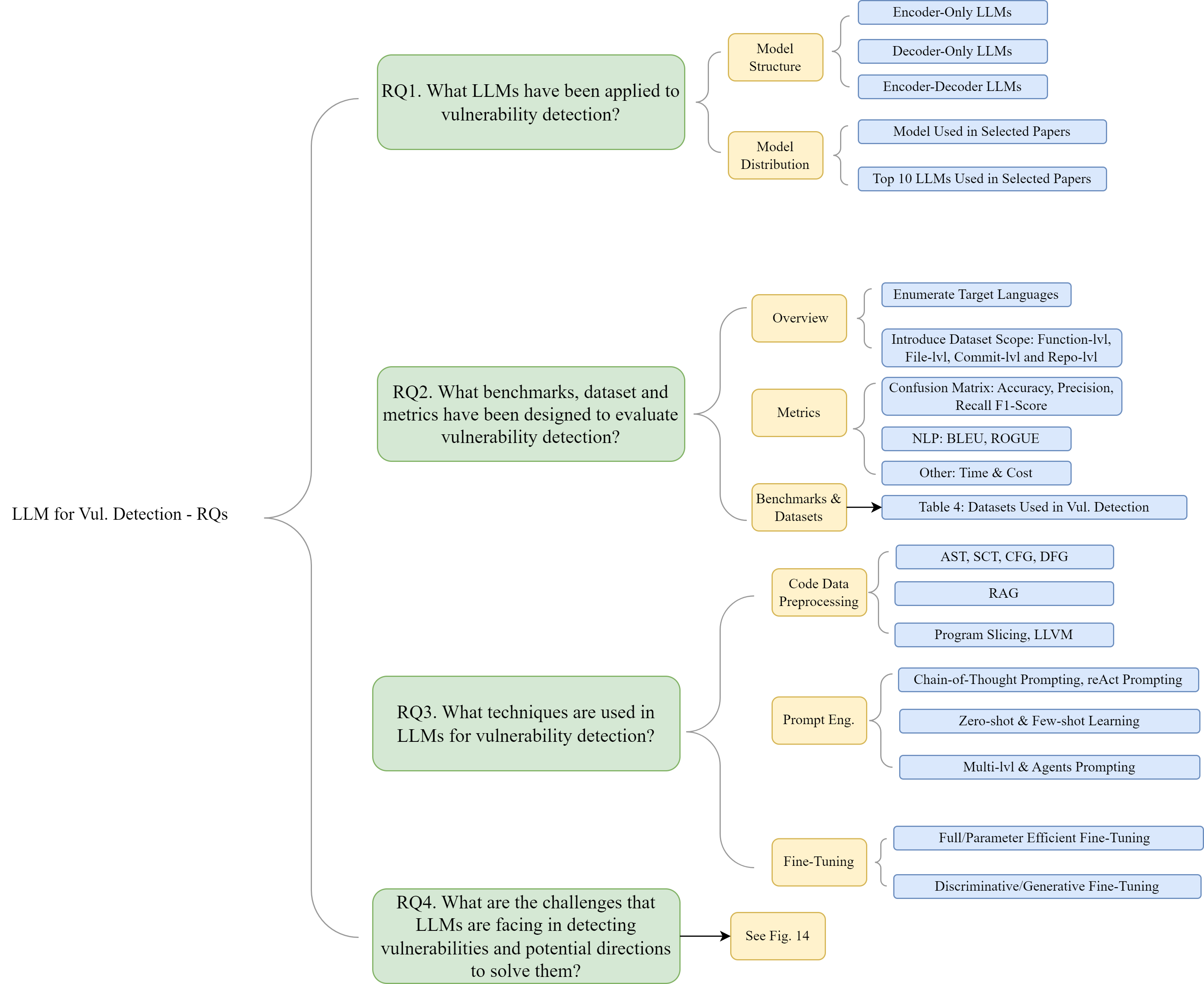}
  \caption{Research Questions}
  \Description{RQs}
  \label{fig:RQs}
\end{figure*}

\section{Research Results}
\subsection{Overview}
Our analysis shows that research on LLM-based vulnerability detection mainly focuses on C/C++, Java, and Solidity. Each language has unique challenges and research priorities. Studies on C/C++ focus on memory-related vulnerabilities, which are critical in this domain. Java research addresses framework-specific vulnerabilities and complex interactions across components in web applications. Solidity research targets vulnerabilities in smart contracts, which are central to blockchain security.

Recently, large decoder-only models, such as GPT and CodeLlama, have become the main choice due to their size and strong generalization abilities. These models are used in 65\% of fine-tuning experiments. For example, Alam et al. \cite{alam2024detection} fine-tuned GPT-4 and achieved 99\% accuracy in Solidity vulnerability detection. Before this, encoder-only models like CodeBERT and GraphCodeBERT were widely used. Advances in prompt engineering have also improved detection performance. Chain-of-Thought prompting is now common for large models and enhances their reasoning abilities for complex code \cite{ding_vulnerability_2024}.

Most current datasets focus on function-level and file-level vulnerabilities, with C/C++ as the dominant target language. Examples include the Devign \cite{zhou2019devign} and CVEFixes \cite{guru2021cvefixes} datasets. However, there is a lack of repository-level datasets that better reflect real-world scenarios. This gap limits the practical use of LLMs in vulnerability detection.

Future research should address challenges in cross-file and complex-context vulnerability detection. It should develop better methods for representing code semantics and create realistic repository-level datasets. These steps will improve the applicability and reliability of LLMs in this field.

\newtcolorbox{finding}[1]{
enhanced,
colback=white,
colframe=gray!75!black,
fonttitle=\bfseries,
title={Finding #1},
colbacktitle=gray!15,
coltitle=black,
attach boxed title to top left={yshift=-2mm, xshift=2mm},
boxed title style={
rounded corners,
size=small,
colframe=gray!75!black
},
boxrule=1pt,
sharp corners,
rounded corners
}

\subsection{RQ1. What LLMs have been applied to vulnerability detection?}
In this paper, we use the LLM categorization and taxonomy outlined by Pan et al. \cite{Pan_2024} and classify the primary LLMs into three architectural groups: 1) encoder-only, 2) encoder-decoder, and 3) decoder-only models. Due to space constraints, we will briefly introduce some representative LLMs in each category. Table~\ref{strengths_weaknesses} gives a clear overview of the strengths and weaknesses of each category of methods based on their inherent capabilities and limitations.

\textbf{Encoder-only LLMs}. Encoder-only LLMs utilize only the encoder component of the Transformer model \cite{Pan_2024}. These models are specifically designed to analyze and represent code or language context without generating output sequences, making them ideal for tasks that demand a detailed understanding of syntax and semantics. By employing attention mechanisms, encoder-only models encode input sequences into structured representations that capture essential syntactic and semantic information \cite{HAUROGNE2024100598}. In the software engineering (SE) domain, encoder-only models such as CodeBERT \cite{feng-etal-2020-codebert}, GraphCodeBERT \cite{guo2021graphcodebertpretrainingcoderepresentations}, CuBERT \cite{10.5555/3524938.3525412}, VulBERTa \cite{Hanif_2022}, CCBERT \cite{zhou2023ccbertselfsupervisedcodechange}, SOBERT \cite{he2024representationlearningstackoverflow}, and BERTOverflow \cite{tabassum2020codenamedentityrecognition} have been widely used.

\textbf{Encoder-Decoder LLMs}.Encoder-decoder models combine both the encoder and decoder components of the Transformer model, allowing them to handle tasks that require both understanding and generation of sequences. The encoder processes an input sequence, transforming it into a structured representation, which is then decoded to produce an output sequence. This structure makes encoder-decoder models versatile for tasks that involve translating, summarizing, or transforming text or code. Prominent examples include PLBART \cite{Ahmad2021UnifiedPF}, T5 \cite{raffel2023exploringlimitstransferlearning}, CodeT5 \cite{wang2021codet5identifierawareunifiedpretrained}, UniXcoder \cite{guo-etal-2022-unixcoder}, and NatGen \cite{10.1145/3540250.3549162}. 

\textbf{Decoder-only LLMs}. Decoder-only LLMs focus exclusively on the decoder component of the Transformer architecture to generate text or code based on input prompts. This approach leverages the model’s capacity to interpret and extend context, enabling it to produce complex and coherent sequences by predicting subsequent tokens. Widely adopted for tasks that emphasize generation, such as vulnerability detection and code suggestion, decoder-only models excel in identifying relevant patterns and potential issues within code. Notable examples in this category include the GPT series (GPT-2 \cite{Radford2019Language}, GPT-3 \cite{NEURIPS2020_1457c0d6}, GPT-3.5 \cite{openai2022gpt35}, GPT-4 \cite{openai2023gpt4}), as well as models tailored specifically for code in software engineering, such as CodeGPT \cite{lu2021codexgluemachinelearningbenchmark}, Codex \cite{chen2021evaluating}, Polycoder \cite{10.1145/3520312.3534862}, Incoder \cite{fried2023incodergenerativemodelcode}, CodeGen series\cite{nijkamp2023codegenopenlargelanguage}, Copilot \cite{github2023copilot}, Code Llama \cite{meta2023codellama}, and StarCoder \cite{li2023starcodersourceyou}. \cite{zhou_large_2024}

\begin{table}[h]
\centering
\caption{Strengths and Weaknesses of Different LLM Categories for Vulnerability Detection}
\label{strengths_weaknesses}
\begin{tabular}{l p{4cm} p{5cm}}
\toprule
\textbf{Category} & \textbf{Strengths} & \textbf{Weaknesses} \\
\midrule
\textbf{Encoder-Only} &
Strong code understanding and static analysis (e.g., CodeBERT) &
Poor at sequence generation and code modification \\
\textbf{Encoder-Decoder} &
Balanced analysis and generation capabilities (e.g., CodeT5) &
High compute cost; lacks task specialization \\
\textbf{Decoder-Only} &
Excels at code generation and patching (e.g., GPT-3.5) &
Limited in contextual understanding and dependency analysis \\
\bottomrule
\end{tabular}
\end{table}

In analyzing 58 studies on vulnerability detection, we identified 33 distinct LLMs used across various tasks. GPT-4 emerged as the most frequently used model, appearing in 29 instances, followed by GPT-3.5 with 25 mentions. Among the categories, encoder-only models represented 24.2\% of total usage, with CodeBERT, GraphCodeBERT, UniXcoder, and BERT as prominent examples. Encoder-decoder models, including CodeT5, made up 8.7\% of usage, serving dual roles in code generation and understanding. Decoder-only models, such as GPT series, CodeLlama, StarCoder, and WizardCoder, accounted for 67.1\% of usage and were primarily applied in code generation tasks.


In addition, Table \ref{top10_llm_used} presents the architecture and occurrences of the top 10 most commonly used LLMs in vulnerability detection research. It shows that most models for this task are decoder-only architectures, indicating a broader usage of this structure in detection tasks. Despite the general trend in the field, which often favors encoder-only architectures for understanding tasks, this table suggests that decoder-only models are also widely adopted for detection, likely due to their efficiency in processing and generating relevant sequences in code analysis.

\begin{table}[h]
\centering
\caption{Top 10 LLMs Used in Vulnerability Detection}
\label{top10_llm_used}
\begin{tabular}{l l l l}
\toprule
\textbf{Usages Ranking} & \textbf{LLM} & \textbf{Structure} & \textbf{Size} \\
\midrule
1  & GPT-4          & Decoder-only    & Unknown \\
2  & GPT-3.5        & Decoder-only    & Unknown \\
3  & BERT           & Encoder-only    & 109M \\
4  & CodeBERT       & Encoder-only    & 125M \\
5  & CodeLlama      & Decoder-only    & 7B, 13B, 34B, 70B \\
6  & LLaMA          & Decoder-only    & 7B, 13B, 70B \\
7  & StarCoder      & Decoder-only    & 15B  \\
8  & CodeT5         & Encoder-Decoder & 220M  \\
9  & Mistral        & Decoder-only    & 7B    \\
10 & GraphCodeBERT  & Encoder-only    & 125M  \\
\bottomrule
\end{tabular}
\end{table}

Among all LLMs, the GPT series (especially GPT-4 series) consistently performs well due to its robust capabilities in understanding and generating code. GPT-4 is widely regarded for advanced applications like vulnerability detection and code analysis, while GPT-3 and GPT-3.5 often serve as baselines or benchmarks in empirical studies. Specialized models such as CodeBERT and CodeT5 are frequently used for fine-tuned tasks involving code understanding and processing. Some research combines multiple models, such as pairing GPT-4 with GPT-3.5, to evaluate comparative performance or execute complementary tasks. This integrated approach, combining general-purpose LLMs with domain-specific models like CodeBERT, leverages the generalization power of LLMs and the task-specific precision of specialized models, resulting in improved performance and versatility.

\newtcolorbox{answerbox}[1]{
    enhanced,
    colback=white,
    colframe=blue!75!black,
    fonttitle=\bfseries,
    title={Answer to #1},
    colbacktitle=blue!15,
    coltitle=blue!75!black,
    attach boxed title to top left={yshift=-2mm, xshift=2mm},
    boxed title style={
        rounded corners,
        size=small,
        colframe=blue!75!black
    },
    boxrule=1pt,
    sharp corners,
    rounded corners
}

\begin{answerbox}{RQ1}
Recent trends show a shift from encoder-only models toward large decoder-only architectures like GPT and CodeLlama series in research. While encoder models still dominate non-fine-tuning studies (72.4\%), decoder-only models account for 65\% of fine-tuning experiments. Encoder-only and encoder-decoder architectures are increasingly positioned as baseline models for comparison.
\end{answerbox}

\subsection{RQ2. What benchmarks, dataset and metrics have been designed to evaluate vulnerability detection?}
In this section, we will start by examining the distribution of vulnerabilities across various programming languages and key software systems. We will then move on to discuss the benchmarks, datasets, and metrics commonly used in the field. Due to differences in design and feature, such as memory management in C/C++, unsafe deserialization in Python, and object injection and reflection in Java, different programming languages have different types of high-occurrence vulnerabilities. This is particularly significant, as many studies on vulnerability detection by LLMs focus on language-specific challenges \cite{Yıldırım2024_evaluating, ferreira2024-smartbugs, li_llm-assisted_2024}. Understanding these nuances is essential to evaluate and improve the effectiveness of vulnerability detection across different contexts.

We analyzed the CVE statistics across major software systems over the past five years (2019-2024), as shown in Table \ref{tab:cve_distribution}.

\begin{figure}[h]
  \centering
  \includegraphics[width=0.35\linewidth]{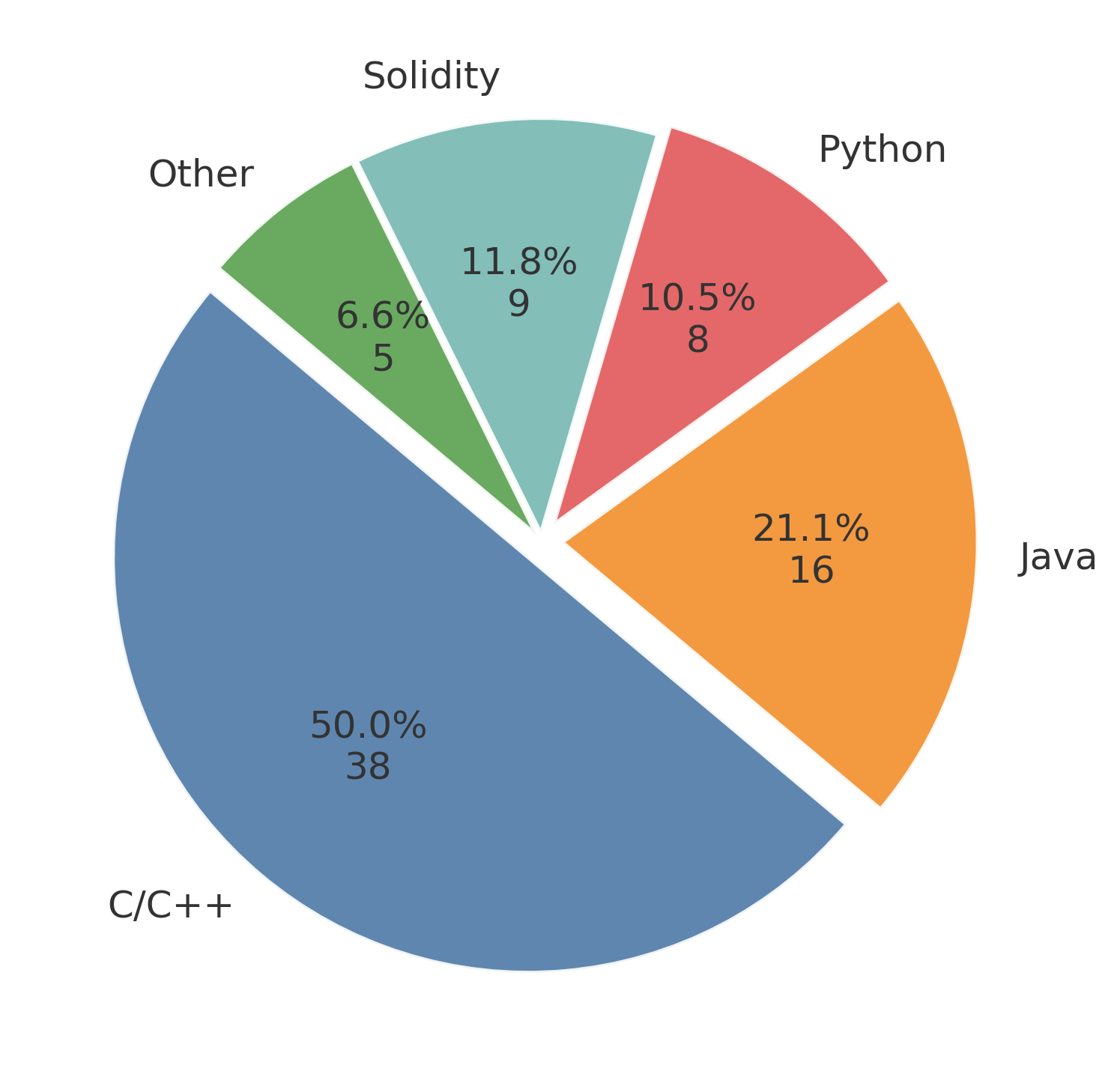}
  \caption{Distribution of Target Programming Languages in LLM-based Vulnerability Detection Research}
  \Description{Target Language Distribution in Selected Papers}
  \label{fig:languages}
\end{figure}

Table \ref{tab:cve_distribution} reveals several interesting patterns in vulnerability distribution across different software systems. Operating systems (Android, MacOS X, Linux Kernel, and Windows Server) dominate the vulnerability landscape, followed by web browsers (Chrome and Firefox) and development platforms (Gitlab). According to CVE statistics \cite{CWE_Metrics}, memory-related vulnerabilities have been the most prevalent type over the past five years. As memory-unsafe yet widely-used programming languages, C/C++ contributes to a significant number of memory corruption vulnerabilities, making vulnerability detection increasingly urgent. Based on our analysis of 56 selected papers, we collected statistics on all target programming languages, as shown in Figure \ref{fig:languages}.

\begin{table*}[h]
\caption{Distribution of CVEs across Different Software Systems}
\label{tab:cve_distribution}
\centering
\begin{tabular}{llccl}
\toprule
\textbf{Software System} & \textbf{Publisher} & \textbf{CVE Number} & \textbf{Primary Language} & \textbf{Software Type} \\
\midrule
Gitlab & Gitlab & 1068 & Ruby & Application \\
Chrome & Google & 3539 & C++ & Browser \\
Firefox & Mozilla & 2700 & C++ & Browser \\
Android & Google & 7215 & Java & Operating System \\
MacOS X & Apple & 3206 & C & Operating System \\
Linux Kernel & Linux & 5912 & C & Operating System \\
Windows Server 2022 & Microsoft & 1607 & C & Operating System \\
\bottomrule
\multicolumn{5}{l}{\small{* Data collected until November 3, 2024 from NVD database}} \\
\end{tabular}
\end{table*}

\begin{finding}{I}
The research landscape shows a clear distribution in target programming languages: C/C++ dominates with 50\% of studies, followed by Java at 21.1\%. Solidity accounts for 11.8\% of research due to its critical role in smart contracts and financial transactions. The remaining 16.6\% covers other languages including Python, PHP, and Go.
\end{finding}

C/C++ remains the primary focus in vulnerability detection, covering 50\% of studies. This high proportion reflects memory-related vulnerabilities common in C/C++ projects. Java ranks second at 21.1\%, partly due to its popularity in enterprise-level software and Android development (Table \ref{tab:cve_distribution} shows Android contributes a large number of CVEs). Java’s type system and bytecode format also provide detailed information for LLMs, and its web applications often face SQL injection, Cross-Site Scripting (XSS), or insecure deserialization. Solidity follows at 11.8\%, as vulnerabilities in smart contracts directly threaten financial security on blockchain platforms. The remaining 16.6\% includes languages like Python, PHP, and Go.

To fine-tune LLMs and measure their performance in vulnerability detection, researchers have introduced various datasets, including BigVul \cite{fan2020bigvul}, CVEfixes \cite{guru2021cvefixes}, and Devign \cite{zhou2019devign}. Each dataset targets different scales, from identifying whether a single function has a vulnerability to scanning an entire GitHub repository. This variation reflects the diverse needs of vulnerability detection tasks.





\textbf{Function-level}. Each data of these datasets contains the following attributes: function implementations (usually including both pre- and post-fix implementations, vulnerable flag (usually 1 for vulnerable and 0 for non-vulnerable). Frequently used datasets of this kind are BigVul \cite{fan2020bigvul} and Devign \cite{zhou2019devign} (also referred to FFmpeg and QEMU dataset). These datasets are often used for fine-tuning large models and evaluating the ability of LLMs to detect vulnerability, but they are not much practical. The reason is that real-world vulnerabilities are usually caused by multiple functions across files.


\begin{figure}[h]
    \centering
    \begin{minipage}[b]{0.45\textwidth}
        \centering
        \includegraphics[width=\textwidth, height=0.25\textheight]{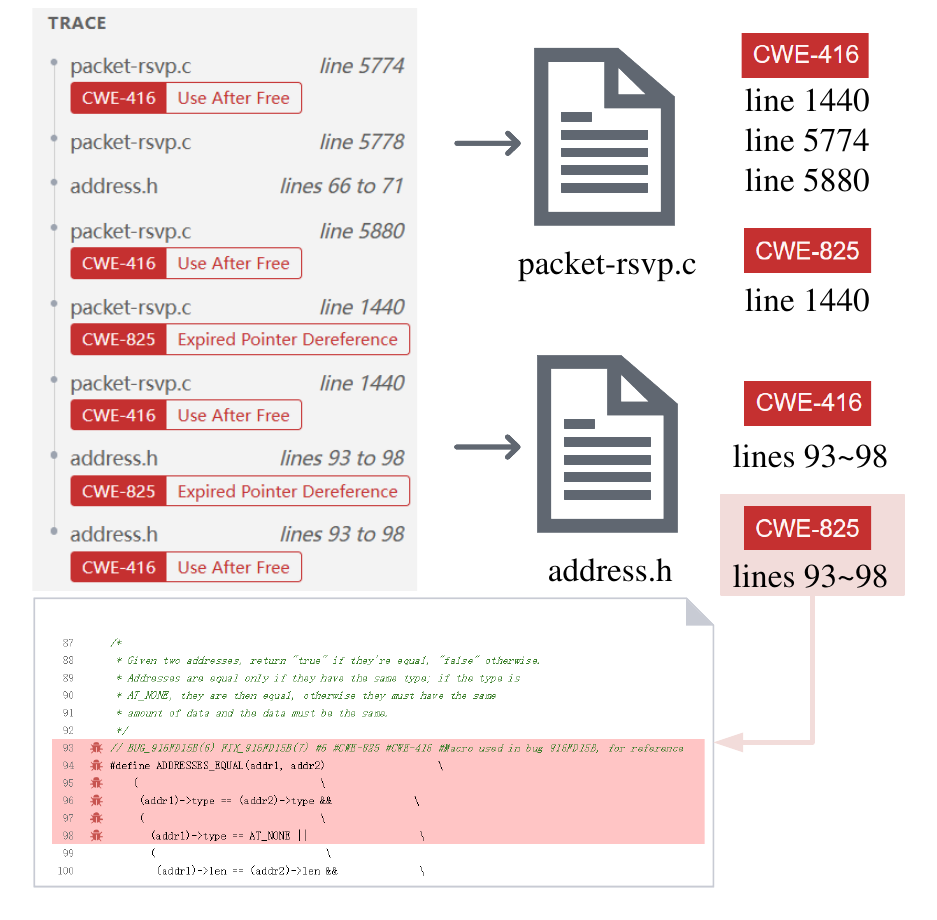}
        \caption{The test case 501129 of Juliet C/C++ 1.3 test suite as a file-level dataset.}
        \Description{A visualization of the file-level dataset structure from test case 501129 of Juliet C/C++ 1.3 test suite}

        \label{fig:file_level}
    \end{minipage}
    \hspace{0.05\textwidth} 
    \begin{minipage}[b]{0.45\textwidth}
        \centering
        \includegraphics[width=\textwidth]{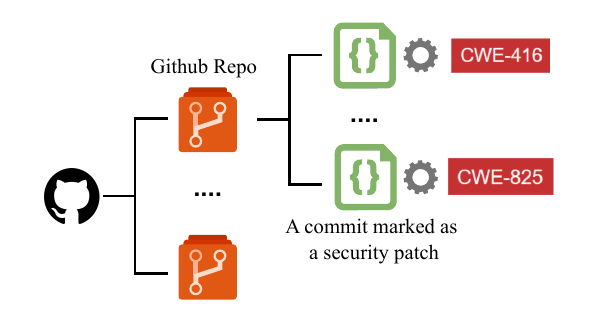}
        \caption{An illustration for a commit-level dataset.}
        \Description{A sample visualization from the commit-level dataset showing the structure and relationships}
        \label{fig:commit_level}
    \end{minipage}
\end{figure}

\textbf{File-level}. Some datasets are structured not only in function level, but also in file level, like Juliet C/C++ \cite{2024nistjulietc} and Java 
\cite{2024nistjulietjava} test suites. Some of the vulnerabilities in the Juliet test suites largely mimic the structure of real-world vulnerabilities, including, but not limited to, cross-file function calls or cross-file access to global variables. Such vulnerabilities with complex contexts provide a significant challenge for LLM to detect vulnerabilities. A test case of Juliet C/C++ test suite has been shown in Figure \ref{fig:file_level}. We can see that in test case 501129, the files and lines where the vulnerabilities are introduced have been marked in the trace. This detail provides clues to LLMs' ability to detect vulnerabilities introduced by multiple files, but also suggests the need to improve LLMs' ability to detect across files.


\textbf{Commit-level}. Many open-source software use the GitHub platform and modify the source code by submitting a commit. Obviously, trusted maintainers can submit commits containing malicious changes that make the target software vulnerable \cite{wu2021feasibility}. So it is also necessary to apply vulnerability detection for each commit. CVEfixes \cite{guru2021cvefixes} and Pan2023 \cite{pan2023} are both commit-level datasets. These datasets typically include the repository URL, commit hash (a unique identifier for each commit), and diff files (showing the differences before and after the commit). In this way, LLMs can analyze the code changes before and after the commit to determine whether the commit is vulnerable or not, and also fetch contextual information through the repository URL.

\textbf{Repository- and application-level}. These types of datasets are usually for vulnerability detection of the entire project. CWE-Bench-Java \cite{li_llm-assisted_2024} is a repository-level datasets focusing on Java projects. Each repository comes with metadata about the vulnerability, such as CWE ID, CVE ID, remediation commits and vulnerability version. This makes analysis and validation more systematic and reliable. And Ghera \cite{mitra2017ghera} is an application-level datasets. Each item contains three applications: a vulnerable application that contains vulnerability X, a malicious application that can attack the vulnerable application using vulnerability X, and a secure application that has no vulnerability X. Each item comes with instructions to build and run the application to demonstrate the vulnerability and its exploitation, thus verifying the presence or absence of the vulnerability and exploitation. An example of commit-level dataset is shown in Figure \ref{fig:commit_level}.

There are also other kinds of datasets. For example, with the rise of blockchain, datasets for smart contract have been built. These datasets, like FELLMVP \cite{luo2024fellmvp}, contains many smart contracts (contract-level) with logical vulnerabilities (e.g. reentrancy attacks and integer overflow/underflow). There are also datasets that focus only on specific vulnerabilities. For example, Code Gadgets \cite{li2018vuldeepecker} only focuses on two types of vulnerabilities in C/C++ programs, buﬀer error vulnerability (CWE-119) and resource management error vulnerability (CWE-399). SolidFi \cite{2020ghaleb-solifi} is only based on injection vulnerability. Frequently used datasets for vulnerability detection are shown in Table \ref{tab:datasets}. The number in parentheses in the size column represents the number of items that are vulnerable.

\begin{table*}
\centering
\caption{Datasets frequently used in LLMs for vulnerability detection}
\begin{adjustbox}{max width=\textwidth}
\begin{threeparttable}
\begin{tabular}{lllcllccc}
\toprule
\textbf{Dataset} & \textbf{Size*} & \textbf{Language} & \textbf{\# Vuln. Types**}  & \textbf{Scope} & \textbf{Source} & \textbf{labeled} & \textbf{Open-source}\\
\midrule
BigVul \cite{fan2020bigvul} & 264,919 (11,823) & C/C++ & 91 & function & real-world & {\color{green}\ding{51}} & \href{https://github.com/ZeoVan/MSR_20_Code_vulnerability_CSV_Dataset}{{\color{green}\ding{51}}} \\
CVEfixes \cite{guru2021cvefixes} & 12,107 (all) & C/C++, Python, PHP & 180 & commit & real-world & {\color{green}\ding{51}} & \href{https://github.com/secureIT-project/CVEfixes}{{\color{green}\ding{51}}}\\
Devign \cite{zhou2019devign} & 27,318 (12,460) & C/C++ & N/A & function & real-world & {\color{green}\ding{51}} & \href{https://github.com/epicosy/devign}{{\color{green}\ding{51}}}\\
Juliet C/C++ \cite{2024nistjulietc} & 64,099 (all) & C/C++ &  118 & file & synthesized & {\color{green}\ding{51}} & \href{https://samate.nist.gov/SARD/test-suites/112}{{\color{green}\ding{51}}}\\
ReVeal \cite{Chakraborty2022reveal} & 18,169 (1,664) & C/C++ & N/A & function & mixed & {\color{green}\ding{51}} & \href{https://github.com/VulDetProject/ReVeal?tab=readme-ov-file}{{\color{green}\ding{51}}}\\
D2A \cite{zheng2021d2a} & 1,295,623 (18,653) & C/C++ & N/A & function & real-world & {\color{green}\ding{51}} & \href{https://github.com/IBM/D2A}{{\color{green}\ding{51}}}\\
SeVc \cite{li2022sevc} & 420,627 (56,395)  & C/C++ & 126 & function & mixed & {\color{green}\ding{51}} & \href{https://github.com/SySeVR/SySeVR}{{\color{green}\ding{51}}}\\
DiverseVul \cite{chen2023diversevul} & 349,437 (18,945) & C/C++ & 150 & function & real-world & {\color{green}\ding{51}} & \href{https://github.com/wagner-group/diversevul?tab=readme-ov-file}{{\color{green}\ding{51}}}\\
SARD \cite{NIST_SARD} & $>$ 5,000,000 (all) & C/C++, Java, PHP & 150 & file & mixed & {\color{green}\ding{51}} & \href{https://samate.nist.gov/SARD/}{{\color{green}\ding{51}}}\\
Juliet Java \cite{2024nistjulietjava} & 28,281 (all) & Java & 112 & file & synthesized & {\color{green}\ding{51}} & \href{https://samate.nist.gov/SARD/test-suites/111}{{\color{green}\ding{51}}}\\
PrimeVul\cite{ding_vulnerability_2024} & 235,768 (6,968) & C/C++ & 140 & function & real-world  & {\color{green}\ding{51}} & \href{https://github.com/DLVulDet/PrimeVul}{{\color{green}\ding{51}}}\\
MAGMA \cite{ahmad2024_magma} & 138 (all) & C/C++, Lua, PHP & 11 & repository & real-world & {\color{green}\ding{51}} & \href{https://hexhive.epfl.ch/magma/}{{\color{green}\ding{51}}}\\
Smartbugs-curated \cite{ferreira2024-smartbugs} & 143 (all) & Solidity & 10 & contract & mixed & {\color{green}\ding{51}} & \href{https://github.com/smartbugs/smartbugs-curated}{{\color{green}\ding{51}}}\\
Smartbugs-wild \cite{ferreira2024-smartbugs} & 47,518 (N/A) & Solidity & N/A & contract & real-world & {\color{red}\ding{55}} & \href{https://github.com/smartbugs/smartbugs-wild}{{\color{green}\ding{51}}}\\
Code Gadgets \cite{li2018vuldeepecker} & 61,638 (17,725) & C/C++ & 2 & function & mixed & {\color{green}\ding{51}} & \href{https://github.com/CGCL-codes/VulDeePecker}{{\color{green}\ding{51}}}\\
VulDeeLocator \cite{li2022_vuldeelocator} & 198,142 (40,450) & LLVM IR & 4 & function & mixed & {\color{green}\ding{51}} & \href{https://github.com/VulDeeLocator/VulDeeLocator}{{\color{green}\ding{51}}}\\
Choi2017 \cite{choi2017_endtoend} & 14,000 ($\approx$ 7,054)  & C/C++ & 4 & function & synthesized & {\color{red}\ding{55}} & \href{https://github.com/mjc92/buffer_overrun_memory_networks}{{\color{green}\ding{51}}}\\
Guo2024	\cite{guo_outside_2024} & 13,532 (6,766)   & C/C++ & N/A & function & real-world & {\color{green}\ding{51}} & \href{https://zenodo.org/records/10975439}{{\color{green}\ding{51}}}\\
SVEN \cite{he2023_large} & 1,606 (all) & C/C++, Python & 9 & commit &  real-world & {\color{green}\ding{51}} & \href{https://github.com/eth-sri/sven}{{\color{green}\ding{51}}}\\
Lin2017	\cite{lin2017} & 6,486 (317) & C/C++ & N/A & function & real-world & {\color{green}\ding{51}} & \href{https://github.com/DanielLin1986/function_representation_learning?tab=readme-ov-file}{{\color{green}\ding{51}}}\\
Ye2024 \cite{ye2024_detecting} & 100 (all) & C/C++ & 1 & application & real-world & {\color{red}\ding{55}} & {\color{red}\ding{55}}\\
ExtractFix \cite{gao2021_beyond} & 7 (all) & C/C++ & 6 & application & real-world & {\color{green}\ding{51}} & \href{https://extractfix.github.io/}{{\color{green}\ding{51}}}\\
DBGbench \cite{marcel2017_how} & 27 (all) & C/C++ & 6 & application & real-world & {\color{green}\ding{51}} & \href{https://dbgbench.github.io/}{{\color{green}\ding{51}}}\\
VulBench \cite{gao_how_2023} & 455 (all) & C/C++, decompiled code & 9 & function & mixed & {\color{green}\ding{51}} & \href{https://github.com/Hustcw/VulBench}{{\color{green}\ding{51}}}\\
VCMatch \cite{wang2022_vcmatch}	& 1,669 (all) & C/C++, Java and PHP & 7 & commit & real-world & {\color{green}\ding{51}} & \href{https://figshare.com/s/0f3ed11f9348e2f3a9f8?file=32403518}{{\color{green}\ding{51}}}\\
Pan2023 \cite{pan2023} & 6,541 (all) & C/C++, PHP, Java & 78 & commit & real-world & {\color{green}\ding{51}} & {{\color{red}\ding{55}}}\\
Ullah2023 \cite{ullah_llms_2024} & 228 (all) &  C/C++, Python & 8 & function & mixed & {\color{green}\ding{51}} & \href{https://github.com/ai4cloudops/SecLLMHolmes}{{\color{green}\ding{51}}}\\
Fang2024 \cite{fang_llm_2024} & 15 (all) & Go, Java, PHP & 9 & application & real-world & {\color{green}\ding{51}} & {{\color{red}\ding{55}}}\\
Ponta2019 \cite{ponta2019_manually} & 1,282 (all) & Java & 6 & commit & real-world & {\color{green}\ding{51}} & \href{https://github.com/SAP/project-kb/tree/main/MSR2019}{{\color{green}\ding{51}}}\\
CWE-Bench-Java \cite{li_llm-assisted_2024}& 120 (all) & Java & 4 & repository & real-world & {\color{green}\ding{51}} & {{\color{red}\ding{55}}}\\
Vulcorpus \cite{kouliaridis_assessing_2024} & 100 (all) & Java & 10 & function & synthesized & {\color{green}\ding{51}} & \href{https://github.com/billkoul/vulcorpus-2024}{{\color{green}\ding{51}}}\\
Vul4j \cite{bui2022_vul4j} & 79 (all) & Java & 25 & commit & real-world & {\color{green}\ding{51}} & \href{https://github.com/tuhh-softsec/vul4j}{{\color{green}\ding{51}}}\\
Ghera \cite{mitra2017ghera} & 69 (all) & Java & 25 & application & synthesized & {\color{green}\ding{51}} & \href{https://secure-it-i.bitbucket.io/ghera/index.html}{{\color{green}\ding{51}}}\\
VjBench	\cite{wu2023_effective}& 42 (all) &  Java & 23 & commit & real-world & {\color{green}\ding{51}} & \href{https://github.com/lin-tan/llm-vul}{{\color{green}\ding{51}}}\\
Yıldırım2024 \cite{Yıldırım2024_evaluating}& 40 (all) & Python & 10 & function & synthesized  & {\color{green}\ding{51}} & {{\color{red}\ding{55}}}\\
FELLMVP \cite{luo2024fellmvp} & 15,637 (820) & Solidity & 8 & contract & real-world & {\color{green}\ding{51}} & \href{https://drive.google.com/drive/folders/1uSXaY7vOvcwQIwXs5JwD9C2hxK9bFMsZ}{{\color{green}\ding{51}}}\\
SolidiFI \cite{2020ghaleb-solifi} & 50 (all) &  Solidity & 7 & contract & real-world & {\color{green}\ding{51}} & \href{https://github.com/DependableSystemsLab/SolidiFI}{{\color{green}\ding{51}}}\\
Ma2024 \cite{ma_combining_2024} & 3,544 (1,734) & Solidity & 5 & function & mixed & {\color{green}\ding{51}} & {{\color{red}\ding{55}}}\\
SC-LOC \cite{zhang2024_empirical}& 1,369 (all) & Solidity & N/A & function & real-world & {\color{green}\ding{51}} & {{\color{red}\ding{55}}}\\
LLM4Vuln \cite{sun_llm4vuln_2024}  & 194 (97) & Java, Solidity & 82 & function & real-world & {\color{green}\ding{51}} & {{\color{red}\ding{55}}}\\
SmartFix \cite{so2023_smartfix}& 361 (all) & Solidity & 5 & contract & mixed & {\color{green}\ding{51}} & {{\color{red}\ding{55}}}\\
Hu2023 \cite{hu_large_2023} & 13 (all) & Solidity & 5 & contract & real-world & {\color{green}\ding{51}} & \href{https://github.com/git-disl/GPTLens}{{\color{green}\ding{51}}}\\
\bottomrule
\end{tabular}
\begin{tablenotes}
\small
\item[*] The number in parentheses represents the number of items that are vulnerable. For example, there are a total of 264,919 functions in the BigVul dataset, of which 11,823 are vulnerable.
\item[**] "N/A" indicates that the authors did not provide detailed information about the number of vulnerabilities in their papers.
\end{tablenotes}
\end{threeparttable}
\end{adjustbox}
\label{tab:datasets}
\end{table*}


\begin{finding}{II}
Current vulnerability datasets exhibit two major limitations: (1) Language imbalance - with C/C++ dominating at around 60\% coverage while Java, despite being widely used in enterprise and Android development, lacks comprehensive datasets; (2) Scope gaps - there is a significant shortage of repository-level datasets that reflect real-world development scenarios where vulnerabilities often span multiple files and dependencies. This scarcity of realistic, large-scale repository datasets poses a critical limitation for practical applications of LLMs in vulnerability detection.
\end{finding}

The evaluation of LLM-based vulnerability detection systems requires multiple metrics. These metrics can be categorized into three groups: classification metrics, generation metrics, and efficiency metrics.

\subsubsection{Evaluation Metrics}
\subsubsection{Classification Metrics}
Vulnerability detection systems commonly use several standard metrics: Accuracy ($\frac{TP + TN}{TP + TN + FP + FN}$) measures overall correctness; Precision ($\frac{TP}{TP + FP}$) indicates true positive rate; Recall ($\frac{TP}{TP + FN}$) shows vulnerability coverage; and F1-Score ($2 \times \frac{Precision \times Recall}{Precision + Recall}$) balances precision and recall. For imbalanced datasets, Matthews Correlation Coefficient (MCC) is also useful:
\begin{equation}
MCC = \frac{TP \times TN - FP \times FN}{\sqrt{(TP + FP)(TP + FN)(TN + FP)(TN + FN)}}.
\end{equation}
\subsubsection{Generation and Explainability Metrics}
To evaluate the quality of generated vulnerability descriptions, Alam et al. and Ghosh et al. \cite{alam2024detection} and \cite{ghosh_cve-llm_2024} employ BLEU and ROUGE metrics. BLEU considers brevity penalty and n-gram precision, while ROUGE measures overlap between generated and reference texts. These metrics help assess both the accuracy and explainability of LLM-based vulnerability detection systems.

\begin{answerbox}{RQ2}
Most benchmarks and datasets for LLM-based vulnerability detection focus on function-level or file-level scope, with C/C++ as the dominant target language.  Classification metrics, like accuracy and precision, are widely used, with Matthews Correlation Coefficient (MCC) adopted for imbalanced datasets. Metrics such as BLEU and ROUGE assess the quality of generated descriptions, while execution time evaluates efficiency. However, current datasets are limited by their focus on C/C++ and lack of repository-level data. These gaps hinder LLMs' ability to generalize across languages and detect complex, multi-file vulnerabilities. Future research should create diverse, large-scale datasets to simulate the real-world scenarios. 
\end{answerbox}

\subsection{RQ3. What techniques are used in LLMs for vulnerability detection?}
Currently, LLM-based vulnerability detection faces several key challenges: (1) data leakage leading to inflated performance metrics, (2) difficulty in understanding complex code context, (3) positional bias in large context windows causing information loss, and (4) high false positive rates and poor performance on zero-day vulnerabilities. Researchers have conducted extensive studies to address these challenges. This section summarizes and discusses current techniques applied to LLM-based vulnerability detection.
\subsubsection{Code Data Preprocessing}
Code processing techniques serve two primary objectives: (1) optimizing the utilization of LLMs' limited context window to improve efficiency, and (2) enhancing LLMs' comprehension of semantic information within the code to improve vulnerability detection capability. 

\begin{figure}[h]
    \centering
    \begin{minipage}[b]{0.45\textwidth}
        \centering
        \includegraphics[width=\textwidth]{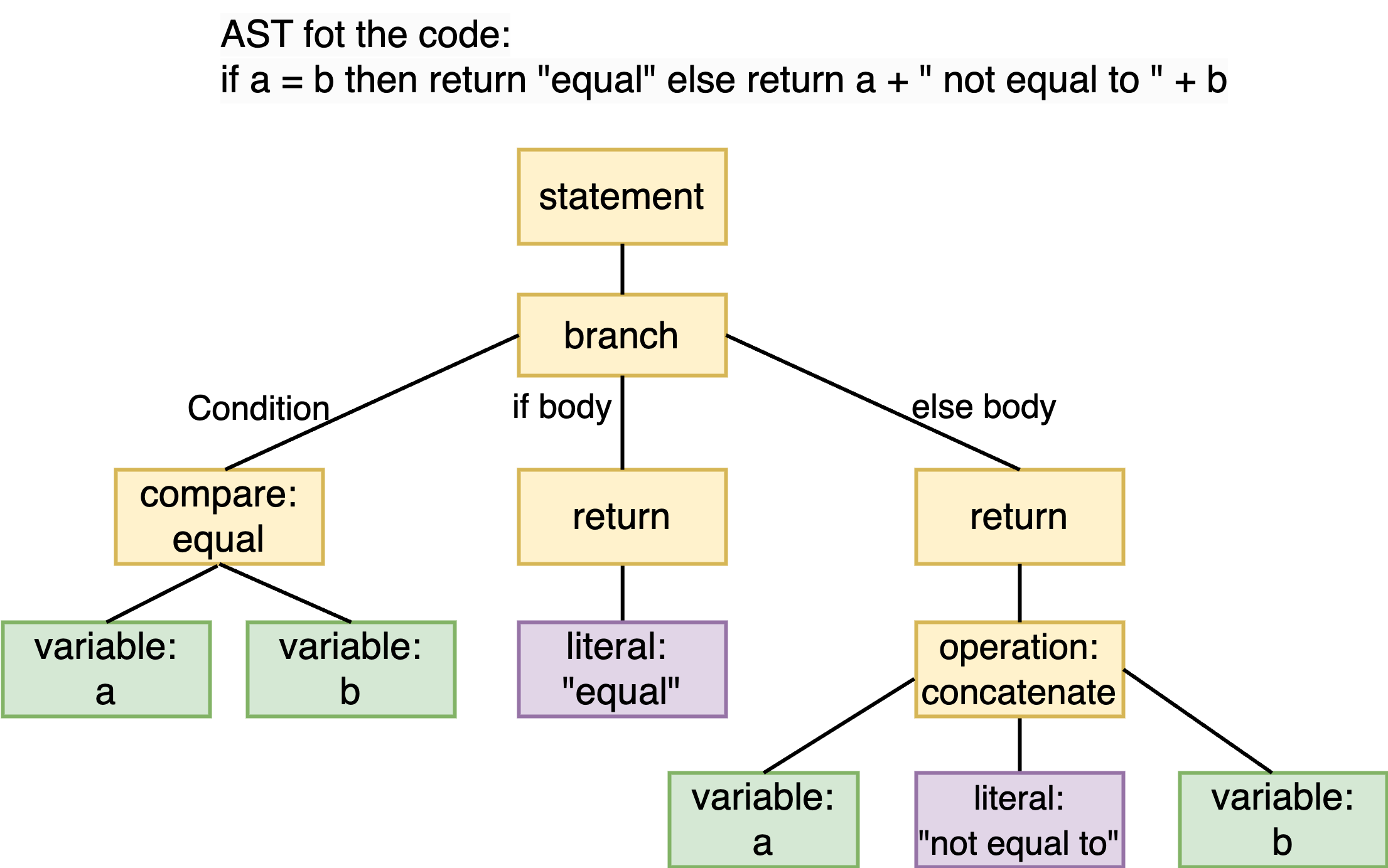}
        \caption{An Example Illustrating AST}
        \label{fig:ast}
    \end{minipage}
    \hspace{0.05\textwidth} 
    \begin{minipage}[b]{0.45\textwidth}
        \centering
        \includegraphics[width=\textwidth]{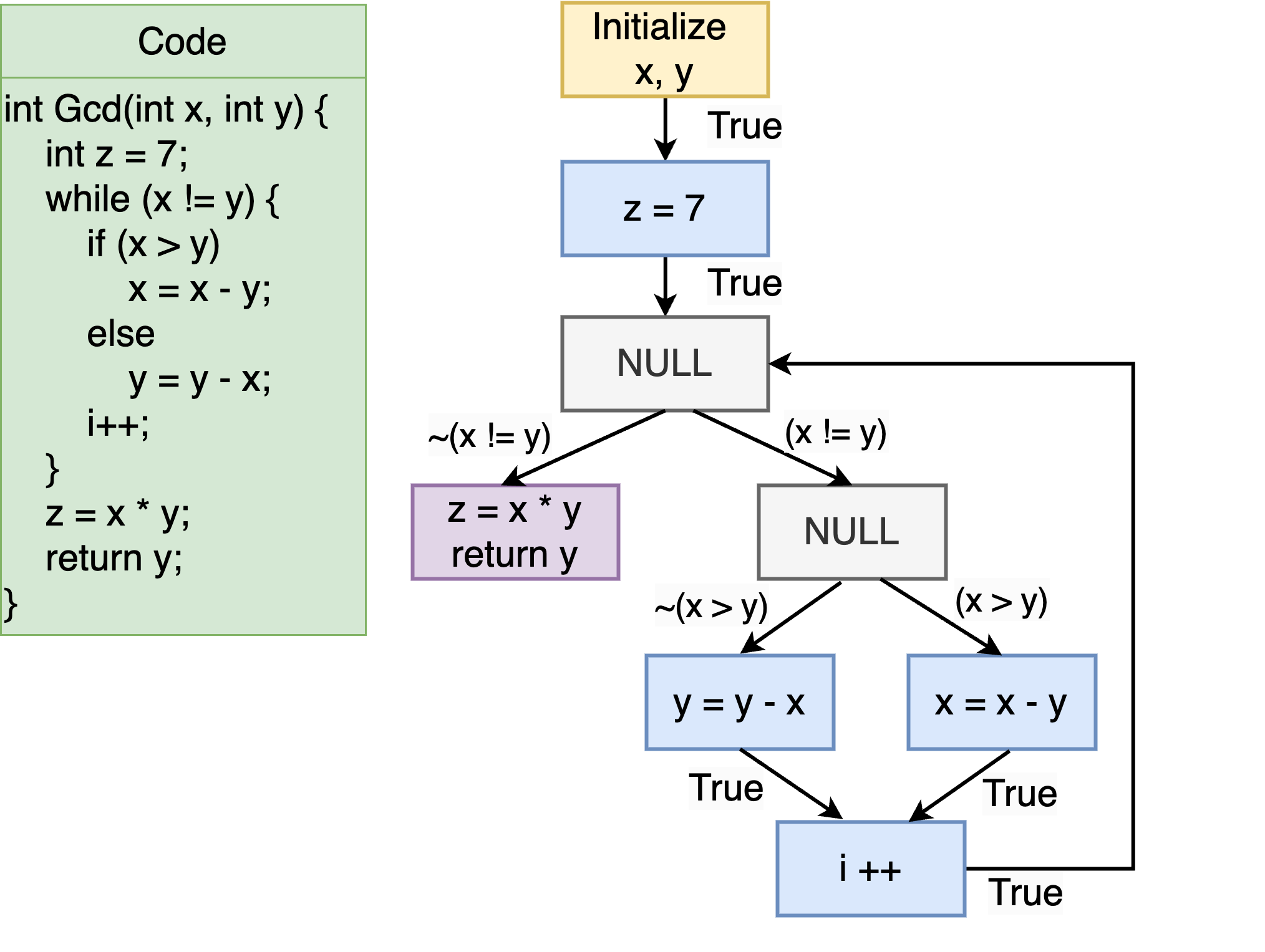}
        \caption{An Example Illustrating DFG}
        \label{fig:dfg}
    \end{minipage}
\end{figure}

\textbf{Abstract Syntax Tree Analysis}. Abstract Syntax Tree (AST) provides a hierarchical representation of program structure, where code elements are organized into a tree format based on their syntactic relationships \cite{welty1997augmenting}. This structural representation eliminates non-essential syntax details while preserving the semantic relationships between code components.  Fig. \ref{fig:ast} represents the AST for a code snippet. AST applications in vulnerability detection can be categorized into several primary approaches: code segmentation and structural representation, where ASTs parse code into function-level segments for efficient processing within LLMs' context limits, as demonstrated by Zhou et al. \cite{zhou_comparison_2024} and Mao et al. \cite{mao_towards_2024}; semantic enhancement, where ASTs are integrated with natural language annotations to form structured comment trees (SCT), as implemented in SCALE framework \cite{wen_scale_2024} to capture vulnerability patterns beyond syntactic relationships; multi-graph analysis, where ASTs are combined with control flow graphs (CFG) and data flow graphs (DFG) to provide comprehensive code structure analysis, as shown in DefectHunter \cite{wang2023defecthunter}; pattern detection integrated with graph attention networks (GATs), exemplified by VulnArmor \cite{sindhwad_vulnarmor_2024}, GRACE \cite{lu_grace_2024}, and \cite{mahyari_harnessing_2024}; and error localization for code evolution as demonstrated in \cite{zhang2024_empirical}. Empirical evaluations across diverse datasets including FFmpeg, QEMU, and Big-Vul demonstrate that AST-augmented approaches significantly improve vulnerability detection performance.


\textbf{Data/Control Flow Analysis}. AST lacks a representation of the data and control flow of a program. Thus, in some papers \cite{tamberg_harnessing_2024, liu_exploration_2024, khare2023understanding, lu_grace_2024, sun_llm4vuln_2024, li_llm-assisted_2024}, data flow graph (DFG) and control flow graph (CFG) have been applied to help LLMs understand the interprocedural data flow and control flow in a program. Fig. \ref{fig:dfg} is a concise example illustrating a Java method and its corresponding DFG. DFG summarizes the possible execution paths in a program, using nodes (or basic blocks, i.e., statements that are executed sequentially without any branching operation) to represent program constructs, and edges to represent the flow of data. CFG has the same basic blocks as nodes, but edges are used to represent the branching operations between basic code blocks. There are two main usages of DFG/CFG like providing extra contextual information in prompt and knowledge base. Combine source code with its DFG and CFG into prompt will result in a significant improvement in LLm's performance in identifying vulnerabilities \cite{tamberg_harnessing_2024, liu_exploration_2024, khare2023understanding, lu_grace_2024}; Sun et al. has proposed that DFG and CFG can be used in knowledge base, with graph-based similarity-search algorithm, to provide LLMs with the information of code segments with similar data and control flow structure \cite{sun_llm4vuln_2024}. Except for DFG and CFG, call graph has been used in vulnerability detection \cite{luo2024fellmvp} to give LLM more information about dependencies between functions.


\textbf{
Retrieval-augmented Generation}. Retrieval Augmented Generation (RAG) enhances the capabilities of LLMs by integrating an information retrieval system that provides extra related information to LLMs \cite{nlprag}. Fig. \ref{fig:ast} illustrates the principle of RAG. LLMs receives user input and applies a searcher to find relevant documents or pieces of information from a knowledge base. The retrieved information is combined with the original prompt to generate a response. In this way, RAG can solve the problem of insufficient knowledge of LLMs in certain domains, illusions and that large language models cannot update data in real time. Many papers have discussed how to choose the right knowledge as a high priority when building RAG for LLMs. \cite{sun_llm4vuln_2024, cao_llm-cloudsec_2024, mathews_llbezpeky_2024, kouliaridis_assessing_2024, keltek_lsast_2024}. Cao et al. directly use CWE database as external knowledge \cite{cao_llm-cloudsec_2024}. Many papers focused on combining code snippets, static analysis results with documentation of corresponding vulnerabilities as knowledge \cite{mathews_llbezpeky_2024, kouliaridis_assessing_2024, keltek_lsast_2024}. In addition to the knowledge mentioned above, Sun et al. used GPT-4 to summarize existing knowledge and thus create two knowledge bases (VectorDB with Vulnerability Report and Summarized Knowledge) \cite{sun_llm4vuln_2024}. RAG has been proven to improve LLM's ability to detect vulnerabilities \cite{keltek_lsast_2024}.


\textbf{Program Slicing}. Program slicing technique has been used to reduce vulnerability-irrelevant lines of code and keep critical lines related to trigger vulnerability \cite{mahyari_harnessing_2024, zhang2024_empirical, purba_software_2023, cao_realvul_2024}. Purba et al. apply program slicing technique to extract code snippets for buffer-overflow detection \cite{purba_software_2023}. These code snippets usually contains key functions, like \texttt{strcmp} and \texttt{memset}, and statements related to call these functions. Cao et al. use program slicing in similar way \cite{cao_realvul_2024}. They first find all potential vulnerability triggers, and apply program slicing technique to collect all statements related to these triggers. While Purba et al. \cite{purba_software_2023} and Cao et al. \cite{cao_realvul_2024} only use the program slicing technique for code preprocess, Zhang et al. proposes to fine-tune LLMs with sliced code to improve the performance of LLMs vulnerability detection \cite{zhang2024_empirical}. Instead of setting explicit slicing criteria, LLMs learns to segregate vulnerable lines of code from a given function during training. This approach helps the model to focus on relevant parts of the code and identify vulnerabilities more accurately. The program slicing technique has been shown to improve LLMs' ability to detect vulnerability \cite{purba_software_2023, zhang2024_empirical}.

\textbf{LLVM Intermediate Representation}.
By converting source code to LLVM Intermediate Representation (IR) \cite{LLVM:CGO04}, analysis and detection methods do not need to be specifically adapted to different programming languages. This improves the versatility of vulnerability detection methods, and LLVM IR also preserves program structure and semantics, making it easier for LLM to analyze potential dependencies in the code. But the downside is obvious: LLVM IR doesn't work with Java and Javascript. To make the approach generalizable to programming languages, the authors converted the source code to LLVM IR and trained LLMs on these IR \cite{mahyari_harnessing_2024}.
%


We can see that in the field of LLMs in vulnerability detection, the techniques mainly comes from traditional software analysis and LLM research. And basically the outputs of these techniques are used as part of the prompts to evaluate the LLMs' vulnerability detection capabilities. These approaches not only optimize the efficiency of LLMs' context window utilization, but also improve its understanding of potential vulnerabilities by preserving or extracting semantic information from the code. However, it also inevitably increases token consumption, and there is also the possibility that too much prompt content may reduce the ability of LLMs to detect vulnerabilities.

\begin{finding}{III}
Our analysis reveals that 41.3\% of studies employed code processing techniques - including graph representations, Retrieval-Augmented Generation (RAG), and code slicing - to better utilize LLMs' limited context windows. While these approaches show modest improvements over direct code prompting, their effectiveness diminishes significantly when dealing with complex, cross-file vulnerabilities. Notably, as larger LLMs (like the GPT series) emerge, the performance gains from the models themselves tend to outweigh those from code processing techniques. This suggests that while current code semantic processing methods offer benefits, developing more effective ways to represent complex code context remains a crucial challenge.
\end{finding}

\subsubsection{Prompt Engineering Techniques}
This is one of the most widely used strategies for optimizing LLM-based vulnerability detection systems, as it enables precise control over model responses by tailoring input prompts.

\begin{figure}[h]
    \centering
    \begin{minipage}[b]{0.45\textwidth}
        \centering
        \includegraphics[width=\textwidth]{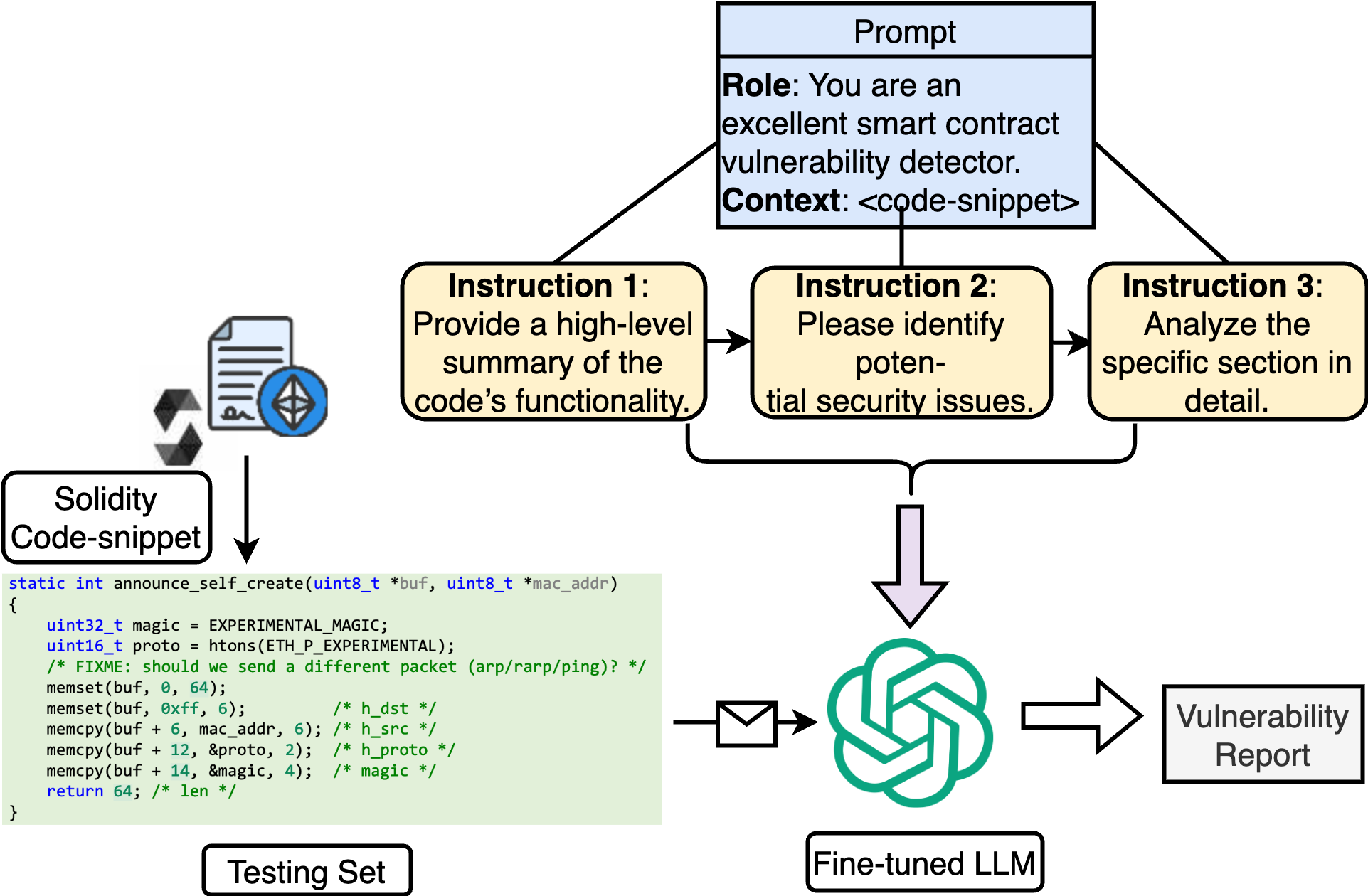}
        \caption{An Instance of Multi-level Prompting for Vulnerability Detection}
        \label{fig:mlevel}
    \end{minipage}
    \hspace{0.05\textwidth} 
    \begin{minipage}[b]{0.45\textwidth}
        \centering
        \includegraphics[width=\textwidth]{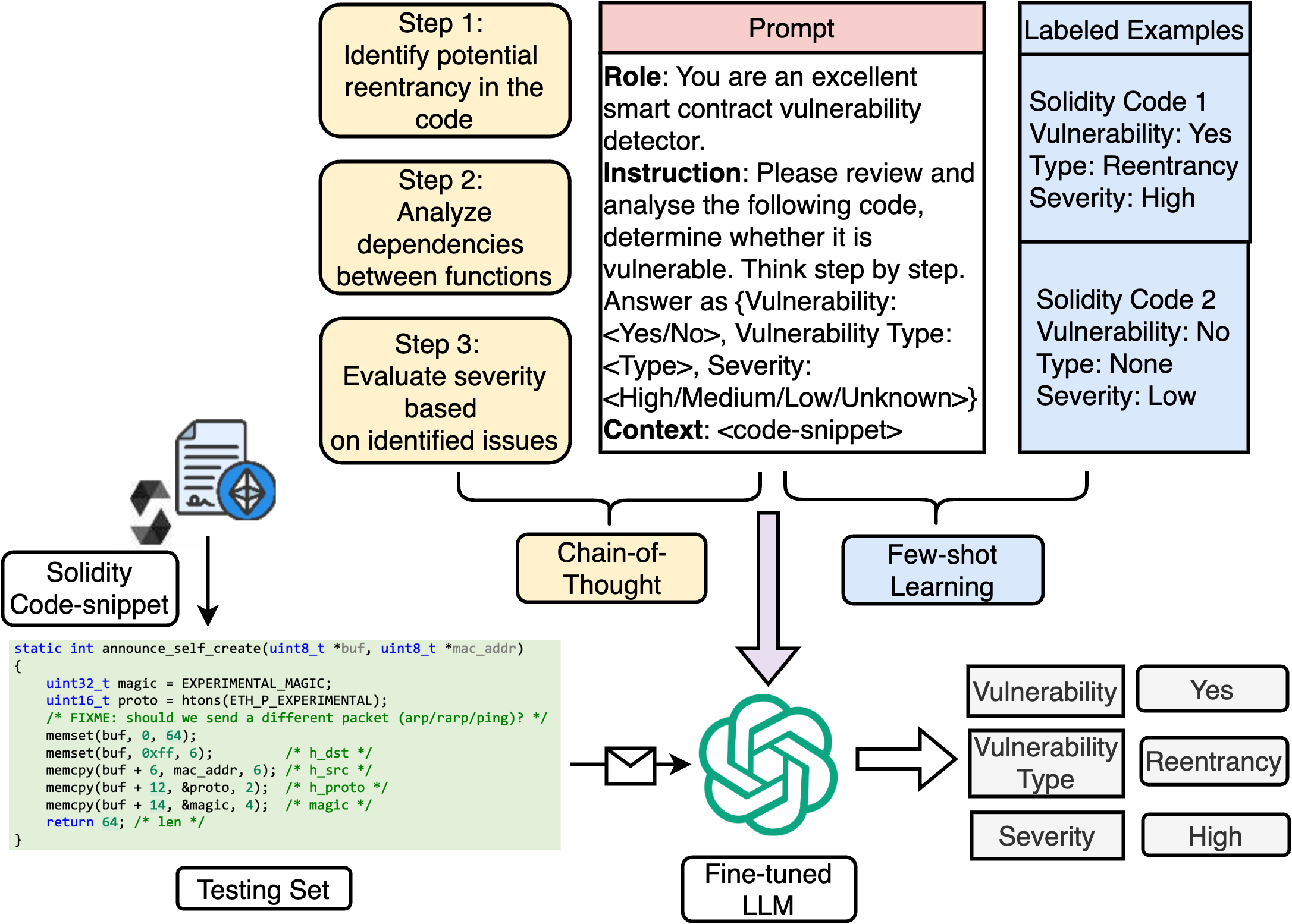}
        \caption{Principles of Chain-of-Thought and Few-shot Learning}
        \label{fig:cot_fsl}
    \end{minipage}
\end{figure}

\textbf{Chain-of-Thought Prompting}. Chain-of-Thought (CoT) Prompting is a technique where LLMs are guided to follow step-by-step instructions to enhance reasoning accuracy before generating a final output. Fig. \ref{fig:cot_fsl} illustrates the COT reasoning process for vulnerability detection. In LLM-based vulnerability detection, CoT prompting involves instructing the model to first summarize the functionality of the given code, then assess potential errors that might introduce vulnerabilities, and finally determine if the code is vulnerable. This structured prompt strategy has been shown to improve precision and recall in vulnerability detection tasks by helping the model reason through complex code in a more organized manner. However, while CoT prompting often enhances precision, its impact on recall can vary across different scenarios \cite{sun_llm4vuln_2024}. 

\textbf{Few-shot Learning}. In LLM-based vulnerability detection, few-shot learning (FSL) enables models to leverage a small set of labeled examples within prompts to improve task-specific performance. In this approach, vulnerability detection can be enhanced by embedding classification standards, such as CWE, directly into the prompt \cite{gao_how_2023}. By incorporating CWE vulnerability categories—complete with numbers and descriptive names—the model gains contextual knowledge that aids in identifying and classifying vulnerabilities accurately. Fig. \ref{fig:cot_fsl} also illustrates the principle of few-shot learning, where the model is provided with labeled examples (e.g., Solidity Code 1 and Solidity Code 2) to understand the task before analyzing a new testing set. These examples, combined with the task-specific prompt, guide the fine-tuned LLM to generate accurate outputs.

\textbf{Hierarchical Context Representation}. Hierarchical context representation is a technique used to manage the context length limitations of LLMs when analyzing extensive codebases. In vulnerability detection, code can be organized hierarchically into modules, classes, functions, and statements. By representing the code in this hierarchical manner, the LLM can process and analyze the code at different levels of abstraction. This approach allows the model to focus on higher-level structures before delving into detailed code segments, effectively managing the context and improving the detection of vulnerabilities within the constraints of LLMs’s maximum input length.

\textbf{Multi-level Prompting}. The multi-level prompting strategy involves breaking down the vulnerability detection task into multiple prompts, each targeting a specific level of analysis. Instead of presenting the entire code and task in a single prompt, the strategy divides the process into stages. For example, the first prompt may ask the LLM to provide a high-level summary of the code's functionality. The second prompt might focus on identifying potential security issues, and subsequent prompts could request detailed analyses of specific sections. This layered approach helps the LLM to systematically process complex code, enhancing its ability to detect vulnerabilities by focusing on one aspect at a time. Fig. \ref{fig:mlevel} illustrates an instance of multi-level prompting.


\textbf{Multiple Prompt Agents and Templates}. This technique employs several specialized prompt agents, each designed with a specific template to perform distinct roles in the vulnerability detection process. For instance, one agent might be tasked with code summarization using a template that guides the LLM to extract key functionalities. Another agent could focus on vulnerability identification, utilizing a template that prompts the model to look for common security weaknesses. By using multiple agents with tailored templates, the system leverages the strengths of each specialized prompt, leading to more accurate and comprehensive vulnerability detection results.

In general, prompt engineering effectiveness varies with model size. Small models benefit from few-shot learning and structured prompting to compensate for limited capabilities, while large models perform better with chain-of-thought prompting and zero-shot approaches due to their stronger generalization abilities and domain knowledge.

\begin{finding}{IV}
As LLMs' inherent capabilities grow, their context window capacity expands accordingly. Chain-of-Thought (CoT) prompting emerges as the dominant approach for large models (>10B parameters), with 100\% of recent studies adopting CoT to enhance generation. For smaller models with limited text processing capacity, zero-shot or minimal few-shot approaches prove more effective, as CoT and extra few-shot examples may cause irrelevant output.
\end{finding}

\begin{table*}
\centering
\caption{Fine-tuning Methods in LLM-based Vulnerability Detection}
\adjustbox{max width=\textwidth}{%
\begin{tabular}{lcccccc}
\toprule
\textbf{Paper} & \textbf{Target Language} & \textbf{FT Method} & \textbf{Main Dataset} & \textbf{Model (F1-Score)} & \textbf{Open-source} \\
\midrule
Alam et al. \cite{alam2024detection} & Solidity & PEFT & VulSmart\cite{alam2024detection} & GPT-4o-mini (0.99) & \color{red}\ding{55} \\
Boi et al. \cite{boi2024smart} & Solidity & PEFT & VulHunt\cite{boi_vulnhunt-gpt_2024} & Llama2-7B-chat-hf & \color{red}\ding{55} \\
Cao et al. \cite{cao_realvul_2024} & PHP & PEFT & RealVul\cite{cao_realvul_2024} & GPT-4 (0.58) & \color{green}\ding{51} \\
Ding et al. \cite{ding_vulnerability_2024} & C/C++ & FFT & PrimeVul\cite{ding_vulnerability_2024} & UnixCoder (0.21) & \color{green}\ding{51} \\
Du et al. \cite{du_generalization-enhanced_2024} & C/C++ & FFT+IFT & Devign\cite{zhou2019devign} & CodeLlama (0.71) & \color{green}\ding{51} \\
Ghosh et al. \cite{ghosh_cve-llm_2024} & C/C++ & DAPT & NVD & MPT-7B (0.93) & \color{red}\ding{55} \\
Gonçalves et al. \cite{gonçalves2024scopeevaluatingllmssoftware} & C/C++ & FFT\&PEFT & CVEFixes\cite{guru2021cvefixes} & NatGen (0.53) & \color{green}\ding{51} \\
Guo et al. \cite{guo_outside_2024} & C/C++ & FFT\&PEFT & Devign\cite{zhou2019devign} & CodeLlama-7bB (0.97) & \color{red}\ding{55} \\
Haurogné et al. \cite{haurogne_advanced_2024} & C/C++ & FFT & DiverseVul\cite{chen2023diversevul} & BERT (0.69) & \color{red}\ding{55} \\
Liu et al. \cite{liu_vuldetectbench_2024} & C/C++ & FFT & SARD \cite{NIST_SARD} & Llama-3-8B & \color{green}\ding{51}  \\
Luo et al. \cite{luo2024fellmvp} & C/C++ & PEFT & Liu2023\cite{liu2023rethinking} & Gemma-7B (0.90) & \color{red}\ding{55} \\
Ma et al. \cite{ma_combining_2024} & Solidity & PEFT & Ma2024\cite{ma_combining_2024} & CodeLlama-13B (0.91) & \color{red}\ding{55} \\
Mao et al. \cite{mao_towards_2024} & C/C++ & PEFT+IFT & SeVC\cite{li2022sevc} & CodeLlama-13B (0.92) & \color{red}\ding{55} \\
Purba et al. \cite{purba_software_2023} & C/C++ & FFT & Code Gadget\cite{li2018vuldeepecker} & Davinci (0.73) & \color{red}\ding{55} \\
Sakaoglu et al. \cite{sakaoglu2023kartal} & HTML/JavaScript & FFT & Sakaolu\cite{sakaoglu2023kartal} & DistilRoBERTa (0.82) & \color{red}\ding{55} \\
Shestov et al. \cite{shestov_finetuning_2024} & Java & PEFT & CVEFixes\cite{guru2021cvefixes} & WizardCoder (0.71) & \color{red}\ding{55}  \\
Taghavi et al. \cite{taghavi_using_2024} & C/C++/Java & PEFT & Mixed Dataset & GPT-4 (0.90) & \color{red}\ding{55}  \\
Wang et al. \cite{wang2023defecthunter} & C/C++ & Model Innovation & QEMU & UnixCoder (0.71) & \color{green}\ding{51}  \\
Wen et al. \cite{wen_scale_2024} & C/C++ & FFT & QEMU & UnixCoder (0.65) & \color{red}\ding{55}  \\
Yang et al. \cite{yang2024large} & C/Java/Python & PEFT & LLMAO\cite{yang2024large} & CodeGen-16B & \color{green}\ding{51}  \\
Yin et al. \cite{yin_multitask-based_2024} & C/C++ & FFT & Big-Vul\cite{fan2020bigvul} & DeepSeek-Coder-6.7B (0.81) & \color{red}\ding{55}  \\
Zhang, C. et al. \cite{zhang2024vtt} & C/C++ & PEFT & Zhang2024\cite{zhang2024vtt} & Llama-7B (0.85) & \color{red}\ding{55} \\
Zhang, J. et al. \cite{zhang2024_empirical} & C/C++/Solidity & FFT & Big-Vul\cite{fan2020bigvul} & CodeLlama-7B (0.82) & \color{red}\ding{55}  \\
Zhou et al. \cite{zhou_comparison_2024} & C/Java/Python & FFT\&PEFT & Zhou2024\cite{zhou_comparison_2024} & Llama-3-8B & \color{red}\ding{55} \\
\bottomrule
\end{tabular}}
\label{tab:finetuning_methods}
\end{table*}

\subsubsection{Fine-tuning}
Fine-tuning helps Large Language Models (LLMs) learn specific tasks better. It works by training pre-trained models again with new data for these tasks.
Fine-tuning is important for three main reasons \cite{zhou2024large, Yao24}: (1) security problems in code follow special patterns that LLMs must learn to find, (2) computer code is different from normal text, so LLMs need to learn how to read and understand code better, and (3) finding security problems needs to be very accurate - missing real problems or reporting false ones can both cause serious issues. As shown in \ref{tab:finetuning_methods}, approximately 30\% of studies choose to fine-tune existing LLMs as their primary proposed method.

\begin{table*}
\centering
\caption{Comparisons of Different Fine-tuning Methods in LLM-based Vulnerability Detection}
\label{tab:ft_comparison}
\begin{tabular}{l p{4cm} p{4cm}}
\toprule
\textbf{Dimension} & \textbf{Method} & \textbf{Properties} \\
\midrule
Parameter Scale & Full Fine-tuning & Updates all parameters for high adaptation capability \\
Parameter Scale & PEFT & Updates subset of parameters for resource efficiency \\
\midrule
Learning Approach & Discriminative & Binary/multi-class classification for precise detection \\
Learning Approach & Generative & Sequence-to-sequence learning with rich output format \\
\bottomrule
\end{tabular}
\end{table*}
\textbf{Full Fine-tuning (FFT). } FFT updates all model parameters during training. Due to computational constraints, most research utilized models with fewer than 15 billion parameters, such as CodeT5, CodeBERT, and UnixCoder. Ding et al.\cite{ding_vulnerability_2024} experimented with five models under 7B parameters, achieving only 0.21 F1-score even when training and validating on PrimeVul. Guo et al.\cite{guo_outside_2024} utilized CodeBERT with FFT for 50 epochs, achieving 0.099 F1-score on PrimeVul but 0.66 F1-score on the Choi2017 dataset. Haurogne et al. \cite{haurogne_advanced_2024} achieved 0.69 F1-score on the DiverseVul dataset, while Purba et al.\cite{purba_software_2023} achieved 0.73 F1-score in buffer overflow detection.

\textbf{Parameter Efficient Fine-tuning (PEFT). } PEFT methods modify only a subset of parameters while keeping most pre-trained weights frozen. Adapters introduce additional trainable layers between original model layers, with Yang et al. \cite{yang2024large} achieving 60\% Top-5 accuracy in fault localization. LoRA represents weight updates as low-rank decompositions, with studies like Du et al. \cite{du_generalization-enhanced_2024} achieving 0.72 F1-score and Guo et al. \cite{guo_outside_2024} reaching 0.97 F1-score on their respective datasets. QLoRA combines parameter quantization with LoRA, as demonstrated by Boi et al.\cite{boi2024smart} achieving 59\% accuracy with lower memory usage.

\textbf{Discriminative Fine-tuning. }For a token sequence \(X = \{x_1, x_2, \ldots, x_L\}\), the model processes it to output vulnerability labels. Zhang et al.~\cite{zhang2024_empirical} and Yin et al.~\cite{yin_multitask-based_2024} demonstrated that a fine-tuned CodeLlama achieves a 0.62 F1-score improvement compared to its non-fine-tuned counterpart.

\textbf{Generative Fine-tuning.} This approach enables sequence-to-sequence learning for generating structured outputs like vulnerability descriptions or vulnerable line identification. Yin et al.\cite{yin_multitask-based_2024} showed that fine-tuning pre-trained LMs outperforms fine-tuning LLMs in generative tasks, with CodeT5+ achieving a ROUGE score of 0.722 compared to DeepSeek-Coder 6.7B's 0.425.

\begin{finding}{V}
Fine-tuning enhances LLM-based vulnerability detection through full and parameter-efficient methods (PEFT). Large models (>10B) with PEFT achieve optimal results, while base models like GPT-4 and CodeLlama deliver F1 scores near 0.9. Discriminative strategies excel in precise detection, requiring datasets with at least 10K samples. However, computational limits and dataset quality remain critical challenges.
\end{finding}

\begin{answerbox}{RQ3}
LLM-based vulnerability detection techniques fall into three categories. First, code preprocessing—such as AST analysis, data/control flow analysis, RAG, and program slicing—enhances context utilization but struggles with complex, cross-file vulnerabilities. Second, prompt engineering—like CoT prompting, few-shot learning, and specialized agents—improves accuracy, with large models (>10B) benefiting from chain-of-thought methods, while smaller models favor simpler prompts. Finally, fine-tuning—both full and parameter-efficient approaches like LoRA—achieves near 0.9 F1-scores, particularly in larger models. As models advance, their inherent strengths may surpass preprocessing benefits, highlighting the need to address complex contexts and cross-file vulnerabilities.
\end{answerbox}


\subsection{RQ4. What are the challenges that LLMs are facing in detecting vulnerabilities and potential directions to solve them?}

\begin{figure*}[h]
   \includegraphics[width=1\linewidth]{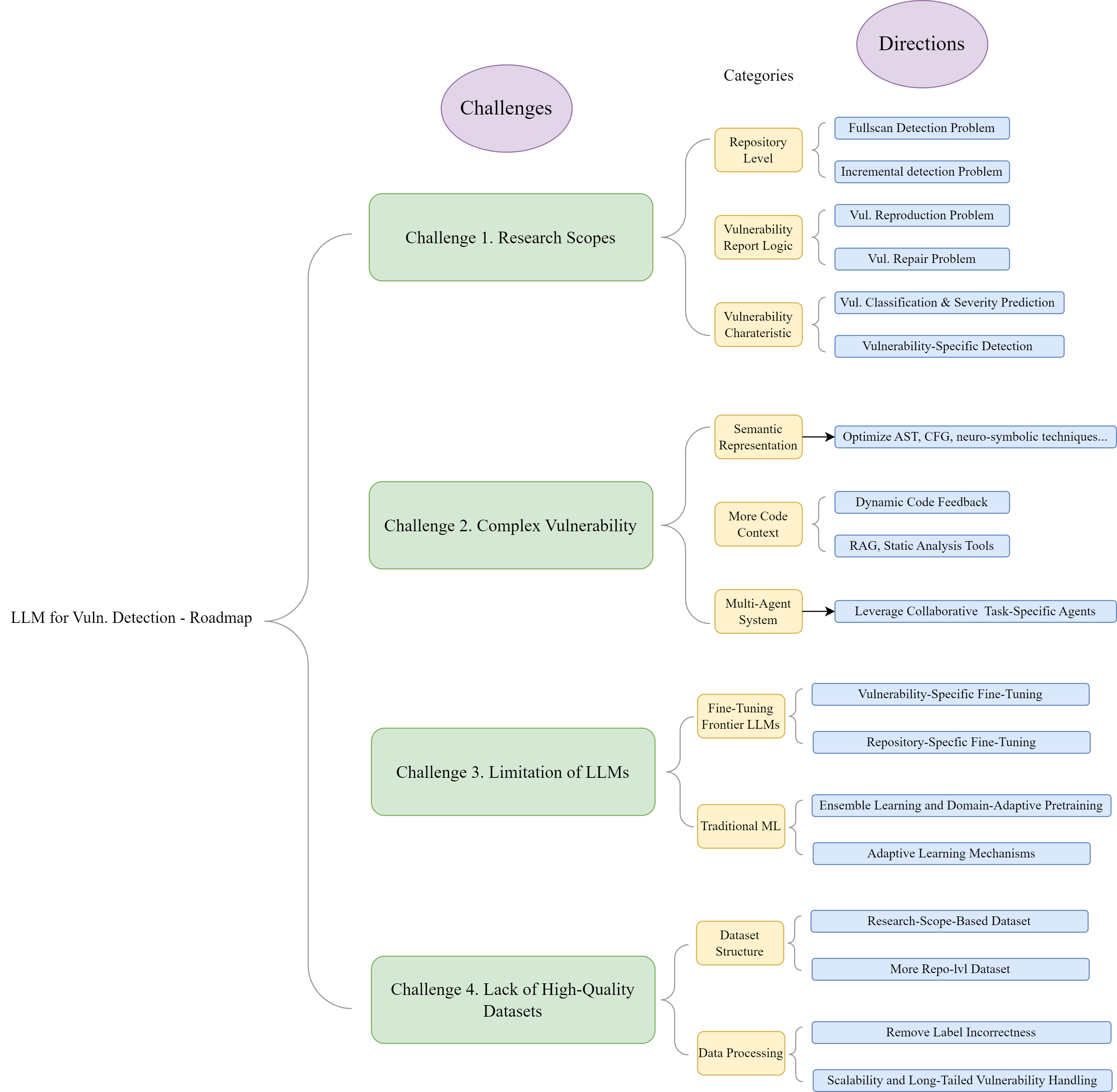}
  \caption{Challenges and Potential Directions}
  \Description{roadmap_rq4}
  \label{fig:roadmap_rq4}
\end{figure*}

The field currently faces four major challenges, along with corresponding potential directions, as illustrated in Figure \ref{fig:roadmap_rq4}. First, researchers struggle to obtain high-quality datasets. Second, large language models (LLMs) show reduced effectiveness when dealing with complex vulnerabilities. Third, these models have limited success in real-world repository applications. Fourth, the models lack robust generation capabilities. Multiple studies confirm these challenges as the main barriers to progress. The following sections examine each challenge in detail.

\textbf{Challenge 1: Limited scope of research problems: } Current research focuses primarily on determining whether a given code snippet is vulnerable or not. In this survey, approximately 40 studies (83\%) concentrate on the analysis of isolated code snippets, where LLM performance often exceeds the results observed in real-world code detection scenarios \cite{gao_how_2023}. While this provides a controlled environment for evaluation, it overlooks the complexities present in practical applications, such as analyzing entire codebases or addressing vulnerabilities that emerge during collaborative development. This indicates that research focused solely on isolated functions or snippets has limited usefulness for real-world scenarios.

\textbf{Potential Directions: }
Beyond analyzing isolated code segments, future research should be structured around the following key problems in real-world development:

Research problems categorized by code evolution in development:
\begin{itemize}

\item \textbf{Full-scale/Incremental Detection:} Full-scale detection requires analyzing entire codebases across multiple files, while incremental detection focuses on new code commits. Current LLMs excel at analyzing isolated code segments but struggle with broader contexts \cite{jiang_when_2024}. As a result, LLMs typically assist static analysis tools or support fuzzing tasks \cite{yang_whitefox_2024, lyu_prompt_2024, yang_kernelgpt_2024} rather than performing standalone analysis. For commit-level detection, existing methods combine commits with static analysis results \cite{li_enhancing_2024, yang_iot_2023}, but may fail when encountering APIs outside their training corpus \cite{jiang_when_2024}.

\end{itemize}

Research problems categorized by vulnerability report workflow:
\begin{itemize}
\item \textbf{Reproducing Vulnerabilities}: Vulnerability reproduction will be essential in future research and a key to reducing false positives. For each detected vulnerability, LLMs should attempt reproduction using input drivers such as fuzzers \cite{jiang_when_2024}. By generating inputs that trigger potential vulnerabilities, LLMs can provide evidence of vulnerability existence \cite{yang_whitefox_2024, xia_fuzz4all_2024}. This approach validates the detection and ensures findings are actionable, thus improving vulnerability report reliability.

\item \textbf{Vulnerability Repair}: While many studies discussed vulnerability repair, practical implementation in real-world projects remains challenging \cite{zhou_large_2024}. Successful vulnerability repair in production environments must meet several criteria:
\begin{itemize}
\item Repaired code must pass all existing tests
\item Repaired code must prevent vulnerability reproduction
\item Repaired code should not introduce new vulnerabilities
\end{itemize}
Current limitations in dataset quality and LLM capabilities hinder effective vulnerability repair validation. While LLMs can identify vulnerable code, they often misidentify vulnerability locations, leading to incorrect explanations and repairs. The ability to generate proof-of-concept exploits and simulate program operations would improve validation, but this requires significant advances in LLM capabilities \cite{kulsum_case_2024}.

\end{itemize}

Research problems categorized by vulnerability characteristics:
\begin{itemize}
\item \textbf{Specialized Vulnerability Detection: }
The development of targeted detection approaches for specific vulnerability categories represents a significant challenge. Current research \cite{gao2021_beyond} demonstrates that LLMs exhibit varying capabilities across different vulnerability types - achieving high accuracy in detecting Out-of-bounds Write vulnerabilities (CWE-787) while performing poorly with Missing Authorization issues (CWE-862). This performance variation necessitates specialized detection mechanisms for distinct CWE categories, particularly for high-frequency vulnerabilities such as memory-related issues. The lack of category-specific research and datasets has left this critical area largely unexplored.

\item \textbf{Vulnerability Classification and Severity Assessment: }
Alam et al. \cite{alam2024detection} and Gao et al. \cite{gao2021_beyond} highlight two fundamental challenges:
First, accurately categorizing detected vulnerabilities according to established frameworks (e.g., CWE) remains difficult. This classification problem is essential to practical development workflows, as different vulnerability types require distinct remediation approaches.
Second, predicting vulnerability severity levels directly influences remediation priorities and timelines, with high-risk vulnerabilities requiring expedited mitigation. 
\end{itemize}

\begin{finding}{VI}
When selecting experimental research problems, researchers should prioritize addressing real-world challenges in software vulnerability detection. For instance, research can be categorized by codebase analysis methods, such as full-scale scanning or incremental scanning, focusing on comprehensive or commit-level vulnerability detection. Additionally, based on the logical workflow of vulnerability reporting in real-world development, research can be divided into vulnerability discovery, reproduction, and repair. Beyond these, other feasible directions include vulnerability classification, severity prediction, and targeted detection for specific vulnerability types.
\end{finding}

\textbf{Challenge 2: Complexity of Representing Vulnerability Semantics.}
Vulnerability patterns are often very complex \cite{li_nuances_2023}. More than 95\% of studies report that code complexity—such as external dependencies, multiple function calls, global variables, and complicated software states—makes it hard to detect vulnerabilities. We can measure this complexity by lines of code or cyclical complexity \cite{tehrani2024assessing}, and visualize it using program dependency graphs (DFGs) or call graphs. However, most methods focus on single code blocks at the function or file level \cite{zhang2024_empirical, du_generalization-enhanced_2024}, which is not very helpful for large projects. When dealing with complex projects, LLMs often have limited input and face much “unseen code” \cite{jiang_when_2024}.

In simpler situations—like a single function of about 500 lines from synthetic datasets—LLMs can detect vulnerabilities well, even in zero-shot settings \cite{cao_realvul_2024, xia_fuzz4all_2024}. But many studies show that more information is needed to handle larger projects \cite{gao_how_2023, ashiwal_llm-based_2024}, especially when the online corpus is sparse. In such cases, we must provide extra documents and specifications \cite{zhang_how_2024}. Also, some functions rely on their callers for protection, so analyzing them alone may lead to incorrect conclusions. We need to give LLMs enough context to identify vulnerabilities accurately.

\textbf{Potential Directions: }
Two core approaches can address this challenge. The first approach focuses on enabling LLMs to read more code. This increases their understanding of the entire repository. The second approach emphasizes using abstract representations to simplify code semantics. This enhances LLMs' comprehension of code structure and behavior.

\begin{itemize}
    \item \textbf{Dynamic Code Knowledge Expansion:} Through feedback loops and adaptive mechanisms, LLMs should be enabled to freely access and understand repository code \cite{yang_kernelgpt_2024, zhang_how_2024}. This would address high false positive rates by providing broader context for vulnerability analysis.

    \item \textbf{Optimized Code Representation:} Studies \cite{tamberg_harnessing_2024, liu_exploration_2024, khare2023understanding, lu_grace_2024, sun_llm4vuln_2024} utilize AST, CFG, and DFG representations to reduce token counts for limited context windows. While current implementations haven't significantly improved detection accuracy, future research opportunities include sophisticated semantic processing, multi-method integration, better context preservation, and hybrid graph representations.

    \item \textbf{Specialized LLM Agents}
    Research explores optimization through specialized LLM agents. Studies \cite{Wang_2024_survey} demonstrate that task division among multiple agents increases output robustness. Each agent specializes in specific aspects of vulnerability detection. Prompt engineering research \cite{liu_exploration_2024} evaluates zero-shot, few-shot, and chain-of-thought approaches for detection accuracy. However, code complexity introduces challenges. Multiple agents show accuracy degradation with complex code. Few-shot and chain-of-thought methods cannot provide sufficient additional context.
    
    \item \textbf{Integration with External Tools}
    External tools provide important support mechanisms. LangChain improves efficiency through simplified and asynchronous LLM calls. Retrieval-Augmented Generation (RAG) gains popularity due to its cost-effectiveness and efficiency. Studies \cite{cao_llm-cloudsec_2024, du_vul-rag_2024} implement RAG to vectorize code contexts and enhance detection through LLM-based retrieval. However, code contexts differ fundamentally from natural language. This difference necessitates specialized approaches for code semantic extraction, storage, and comparison.

These optimization techniques could potentially bridge the gap between LLMs' current capabilities and the complex requirements of real-world vulnerability detection in large codebases.
\end{itemize}

\begin{finding}{VII}
The complexity of vulnerabilities indicates that vulnerable code often involves intricate control flows. Addressing this requires improving LLMs' ability to efficiently store and process code information. Researchers can use retrieval-augmented generation (RAG) or similar tools to dynamically expand knowledge. Code semantics can be compressed using control flow graphs (CFGs), abstract syntax trees (ASTs), or neural-symbolic methods. Additionally, specialized agents optimized for specific tasks can be employed within vulnerability detection systems. These approaches enhance the efficiency and effectiveness of LLMs in handling complex code structures.
\end{finding}

\textbf{Challenge 3: Intrinsic Limitations of LLMs.}
Detection solutions must maintain robustness against data changes and adversarial attacks \cite{du_generalization-enhanced_2024}. However, research by Yin et al. \cite{yin_multitask-based_2024} reveals that LLMs lack this robustness. They show vulnerability to data perturbations. These findings indicate the need for more reliable approaches.

Additionally, LLMs need better explainability and consistency for real-world applications. Haurogne et al. and Du et al. \cite{haurogne_advanced_2024, du_generalization-enhanced_2024} demonstrate that LLMs produce inconsistent vulnerability explanations. They cannot guarantee correct explanations in every instance. The outputs show randomness across different runs. Even when LLMs correctly identify vulnerable code, they often fail to provide accurate vulnerability explanations. This limitation creates significant problems for subsequent repair and review processes. Current research in this area remains insufficient.


Future research should focus on these key areas: improving accuracy, enhancing robustness, and increasing output reliability and explainability. These improvements will make LLM-based solutions more practical for real-world use.

\textbf{Potential Directions: }
The core to this challenge is to improve the inherent vulnerability detection ability of LLMs. Researchers may focus on training new models or fine-tuning LLMs.
\begin{itemize}
    \item \textbf{Fine-tune Frontier LLMs: }  
    Recent findings on scaling laws \cite{kaplan2020scalinglawsneurallanguage} indicate that larger decoder-only language models, such as GPT-4 and Claude3.5-Sonnet, can achieve systematically improved performance as model size, training data, and compute are scaled up. Improvements in model capabilities enhance vulnerability detection, classification, and explanation. Research by Alam et al. \cite{alam2024detection} shows that GPT-4 and GPT-3.5 achieve higher detection accuracy than earlier models like Llama2 and CodeT5 under identical prompts. The release of GPT-O1, with its visible reasoning process, suggests improvements in both detection capability and output explainability.
    
    Fine-tuning approaches show promise. Several studies \cite{alam2024detection, cao_realvul_2024, du_generalization-enhanced_2024, guo_outside_2024, mao_multi-role_2024, ma_combining_2024} fine-tune open-source models like CodeLlama and CodeBERT. These achieve results comparable to general LLMs. However, dataset limitations present challenges. Research \cite{vieira2024much} indicates that practical applications require datasets of at least 100,000 examples. This creates significant training cost barriers. Researchers can enhance vulnerability detection performance through the following approaches:
    
    \begin{itemize}
        \item Vulnerability-specific Fine-tuning: Fine-tune models on specific vulnerability types. Research \cite{alam2024detection, cao_realvul_2024, du_generalization-enhanced_2024} shows models trained on specific vulnerability categories (e.g., memory-related issues, injection flaws) achieve higher detection accuracy. This targeted approach allows models to learn deeper patterns within each vulnerability type.
    
        \item Repository-adaptive Fine-tuning: Adapt models to specific codebases through fine-tuning on repository-specific data. Studies \cite{Hanif_2022, mahyari_harnessing_2024} demonstrate this approach improves detection accuracy by helping models understand project-specific coding patterns and architecture. This method benefits large, complex projects with unique coding conventions.
    \end{itemize}
    
\end{itemize}

\begin{itemize}
   \item \textbf{Ensemble Learning and Domain-Adaptive Pretraining:} Combining predictions from multiple models effectively reduces false positives and improves detection accuracy. DAPT can refine LLMs' understanding of specific contexts by leveraging curated datasets including both public records (e.g., NVD) and domain-specific data. This enables better identification of niche vulnerabilities and improved generalization.

   \item \textbf{Adaptive Learning Mechanisms:} To address the dynamic nature of security threats and enhance model robustness, adaptive learning \cite{martin2020systematic} mechanisms allow continuous knowledge updates through feedback loops and periodic retraining. Advanced optimization techniques can further improve prompt configurations and learning rates, ensuring reliability in real-world applications.
\end{itemize}

\begin{finding}{VIII}
Enhancing robustness and explainability in LLMs is essential for effective vulnerability detection. Fine-tuning on specific vulnerability types, such as memory issues or injection flaws, improves detection accuracy by focusing on targeted patterns. Repository-adaptive fine-tuning helps models learn project-specific coding conventions, further increasing accuracy. Ensemble learning combines predictions from multiple models to reduce false positives, while domain-adaptive pretraining (DAPT) refines model understanding of niche contexts using curated datasets. Adaptive learning mechanisms, incorporating feedback loops and periodic updates, ensure LLMs remain robust against evolving threats. These methods address LLM limitations and improve their real-world applicability.
\end{finding}

\textbf{Challenge 4: Lack of High-Quality Datasets.}
High-quality vulnerability benchmark datasets remain scarce. Current datasets face several problems. These include data leakage, incorrect labels, small size, and limited scope \cite{mathews_llbezpeky_2024, wu2023_effective, chen2023diversevul}. 

\textbf{Dataset Incorrectness:}
A critical issue in these challenges is the incorrect labeling of vulnerabilities, which harms the reliability and effectiveness of datasets. Automated collection methods \cite{nie23Understanding} can gather large amounts of data quickly but cannot ensure correct labels without human review, leading to many mislabeled or inaccurately annotated samples. Research \cite{wu2023_effective} also shows frequent data leakage, as LLMs often train on sources like GitHub, old software versions, and external libraries with inadequate version control or deduplication. As a result, models may encounter the same test data seen during training, inflating performance metrics and undermining real-world validity. In conclusion, a high-quality dataset for vulnerability detection should meet several requirements.

\begin{itemize}
    \item \textbf{Accurate Labels.} Since labels are crucial in supervised learning, incorrect annotations can lead to serious issues in production environments.
    \item \textbf{Minimal Data Leakage.} Large-scale LLMs trained on broad codebases risk seeing identical vulnerabilities during testing. Countermeasures include code obfuscation, synthesis, and using updated datasets.
    \item \textbf{Comprehensive Annotations.} For repository-level data, providing call sequences and control flows that reproduce the vulnerability, as well as detailed descriptions, helps LLMs create more reliable detection reports.
\end{itemize}

\textbf{Potential Directions: }
The key to addressing this challenge lies in researchers' focus on constructing datasets based on specific research scopes. As discussed in Challenge 1, different research problems in LLM-based vulnerability detection require distinct types of datasets, all of which currently lack sufficient accurate data or cases. Researchers should approach dataset construction with targeted focus on specific research problems. Researches can be developed on:

\begin{itemize}
    \item \textbf{Dataset Quality and Scope Enhancement.}
    Research can focus on developing smaller, high-quality test sets to effectively measure progress in vulnerability detection. One approach combines existing verified samples from multiple studies \cite{alam2024detection, zibaeirad2024vulnllmevalframeworkevaluatinglarge, liu_vuldetectbench_2024}. This creates a reliable test benchmark that the research community can maintain and expand over time. Moreover, recent advances in LLMs, particularly GPT-4o with its 128k context window, enable comprehensive repository-level vulnerability analysis, allowing researchers to detect and repair vulnerabilities across entire codebases rather than just at function-level.

    \item \textbf{Scalability and Long-Tailed Vulnerability Handling:} Handling long-tailed distributions of vulnerability types requires both scalable models and data augmentation techniques \cite{liu_enhancing, deng_improving}. Generating synthetic samples for rare vulnerability types can improve LLMs’ ability to detect low-frequency events. Integrating structured information, such as CWE classifications, can further enhance the model's capability to prioritize and address critical vulnerabilities effectively \cite{atiiq2024generalist}.
\end{itemize}

\begin{finding}{IX}
High-quality datasets are essential for advancing LLM-based vulnerability detection. Repository-level datasets with detailed annotations, including call sequences and control flows, enhance real-world applicability. Targeted datasets aligned with specific research scopes address distinct detection challenges. Synthetic data generation mitigates data leakage and handles rare vulnerability types. Combining verified samples with scalable data augmentation ensures robust benchmarks for repository-wide vulnerability detection.
\end{finding}

\begin{answerbox}{RQ4}
The main challenges in LLM-based vulnerability detection include research scopes, dataset quality, vulnerability complexity, and model robustness. Key research directions involve improving model capabilities, developing advanced usage methods, enhancing datasets, strengthening detection robustness, and specializing vulnerability detection approaches. There is still a long way to go.
\end{answerbox}

\section{Limitations}
Several factors may affect this survey's comprehensiveness. First, approximately 60\% of research in LLM-based vulnerability detection appears as preprints on arXiv. This reflects the field's emerging nature. Second, terminology variations in concepts like "LLM" and "vulnerability detection" may lead to oversights in initial searches. To mitigate these risks, we implemented a systematic approach. We began by analyzing published papers from established conferences and journals. We extracted core keywords from these sources. Over a two-month period, we refined our selection from approximately 500 papers to 58 highly relevant studies. Future versions of this survey will incorporate new developments in this rapidly evolving field. This ongoing process will ensure more comprehensive and timely coverage of research literature.

\section{Conclusion}
This study presents a systematic analysis of LLM applications in vulnerability detection. Through extensive literature review, we provide a comprehensive examination of the current research landscape, systematically addressing four key questions: the application of LLMs in vulnerability detection, the design of evaluation benchmarks and datasets, current technical approaches, and existing challenges with future directions.

Our findings demonstrate that LLMs exhibit significant potential in code comprehension and vulnerability detection. Through techniques such as fine-tuning and prompt engineering, LLMs can effectively improve detection accuracy. Experiments across multiple benchmark datasets indicate that recent large-scale LLMs, such as GPT-4 and Claude-3.5, have achieved notable progress in vulnerability detection tasks.
However, significant challenges remain in applying LLMs to practical security development. The primary obstacle is the scarcity of high-quality datasets, which constrains model training and evaluation. Additionally, current LLMs show notable limitations in handling complex code structures and repository-level vulnerability detection. Furthermore, issues regarding output randomness and model explainability require further investigation.

Based on these findings, we propose several promising research directions: enhancing model adaptation to code evolution, improving vulnerability reproduction and repair capabilities, developing high-quality datasets, and strengthening model robustness and explainability. Advances in these areas will drive the broader adoption of LLMs in vulnerability detection.

In the future, we plan to enrich this review by adding more vulnerability-related tasks, such as vulnerability localization, vulnerability assessment and vulnerability patching.



\appendix

\bibliographystyle{ACM-Reference-Format}
\bibliography{main}










\end{document}